\documentclass[12pt]{article}
\usepackage{epsfig}
\usepackage{psfrag}
\usepackage{latexsym}
\usepackage{indentfirst}
\usepackage{fancyhdr}
\usepackage{amssymb}
\usepackage{amsmath}
\usepackage{amsfonts}
\usepackage{cite}
\usepackage{bbold}
\usepackage{color}
\usepackage[footnotesize]{caption2}
\usepackage{graphicx}
\usepackage[center,footnotesize,hang]{subfigure}
\usepackage{url}

\def\be{\begin{equation}}
\def\ee{\end{equation}}
\def\bc{\begin{center}}
\def\ec{\end{center}}
\def\bea{\begin{eqnarray}}
\def\eea{\end{eqnarray}}

\catcode`@=11
\def\marginnote#1{}
\newcount\hour
\newcount\minute
\newtoks\amorpm
\hour=\time\divide\hour by60
\minute=\time{\multiply\hour by60 \global\advance\minute by-\hour}
\edef\standardtime{{\ifnum\hour<12 \global\amorpm={am}%
        \else\global\amorpm={pm}\advance\hour by-12 \fi
        \ifnum\hour=0 \hour=12 \fi
        \number\hour:\ifnum\minute<10 0\fi\number\minute\the\amorpm}}
\edef\militarytime{\number\hour:\ifnum\minute<10 0\fi\number\minute}
\def\draftlabel#1{{\@bsphack\if@filesw {\let\thepage\relax
   \xdef\@gtempa{\write\@auxout{\string
      \newlabel{#1}{{\@currentlabel}{\thepage}}}}}\@gtempa
   \if@nobreak \ifvmode\nobreak\fi\fi\fi\@esphack}
        \gdef\@eqnlabel{#1}}
\def\@eqnlabel{}
\def\@vacuum{}
\def\draftmarginnote#1{\marginpar{\raggedright\scriptsize\tt#1}}
\def\draft{\oddsidemargin 0.0truein
        \def\@oddfoot{\sl preliminary draft \hfil
        \rm\thepage\hfil\sl\today\quad\militarytime}
        \let\@evenfoot\@oddfoot \overfullrule 3pt
        \let\label=\draftlabel
        \let\marginnote=\draftmarginnote
   \def\@eqnnum{(\theequation)\rlap{\kern\marginparsep\tt\@eqnlabel}%
\global\let\@eqnlabel\@vacuum}  }
\catcode`@=12
\begin{document}
\begin{titlepage}
\vspace*{-1cm}
\phantom{hep-ph/***}


\vskip 2.5cm
\begin{center}
{\Large\bf Tri-Bimaximal Neutrino Mixing and the $T_{13}$ Flavor
Symmetry}
\end{center}
\vskip 0.2  cm
\vskip 0.5  cm
\begin{center}
{\large Gui-Jun Ding}~\footnote{Email: dinggj@ustc.edu.cn}
\\
\vskip .2cm {\it Department of Modern Physics,}
\\
{\it University of Science and Technology of China, Hefei, Anhui
230026, China}
 \vskip .2cm
{\it Department of Physics, University of
Wisconsin-Madison,}\\
{\it 1150 University Avenue, Madison, WI 53706, USA}
\end{center}
\vskip 0.7cm
\begin{abstract}
\noindent

We present a supersymmetric $T_{13}$ model for the tri-bimaximal
neutrino mixing, and the complete flavor group is $T_{13}\times
Z_{4}\times Z_2$. At leading order, the residual symmetry of the
charged lepton sector is $Z_3$, and the $T_{13}$ symmetry is broken
completely in the neutrino sector. The charged lepton mass
hierarchies are determined by the spontaneous breaking of the flavor
symmetry, both the type I see-saw mechanism and the Weinberg
operator contribute to generating the light neutrino masses.
Tri-bimaximal mixing is exact at leading order while subleading
contributions introduce corrections of order $\lambda^2_c$ to the
three lepton mixing angles. The vacuum alignment and subleading
corrections are studied in detail, a moderate hierarchy of order
$\lambda_c$ between the vacuum expectation values of the flavon
fields in the charged lepton and neutrino sectors can be
accommodated.

\end{abstract}

\end{titlepage}
\setcounter{footnote}{1}
 \vskip2truecm

\section{Introduction}
The presence of two large and one small mixing angles in the lepton
sector \cite{Schwetz:2008er,Fogli:Indication,GonzalezGarcia:2010er},
\begin{equation}
0.27<\sin^2\theta_{12}<0.37,~~~~0.39<\sin^2\theta_{23}<0.64,~~~~
\sin^2\theta_{13}<0.040\,(0.044)~~~{\rm at}~3\sigma
\end{equation}
suggests that the observed neutrino mixing matrix is remarkably
compatible with the so called tri-bimaximal (TB) structure
\cite{TBmix} within measurement errors. The simple form of the TB
mixing matrix implies an underlying family symmetry between the
three generations of leptons. It has been realized that the TB mixing
can naturally arise as the result of a particular vacuum alignment
of scalars that break spontaneously certain flavor symmetries. In
the past years, much effort has been devoted to produce TB mixing
based on some family symmetry. It has been shown that TB mixing can
be understood with the help of discrete flavor symmetries, such as
$A_4$ \cite{TBModel,Csaki:2008qq,Altarelli:2008bg,Morisi:2007ft,Feruglio:2008ht,Hagedorn:2009jy}, $T_7$
\cite{T7} , $T^{\prime}$ \cite{Tprime}, $S_4$
\cite{S4model,Lam:2008rs} and $\Delta(27)$ \cite{Delta27}, or
continuous flavor symmetry $SO(3)$ \cite{SO(3)} and $SU(3)$
\cite{SU(3)}. Discrete non-abelian groups appear to be particularly
suitable to reproduce the TB mixing pattern, some higher order
discrete groups such as $A_5$ \cite{Everett:2008et}, $\Delta(54)$
\cite{Escobar:2011mq}, $\Sigma(81)$ \cite{Ma:2006ht} and $PSL_2(7)$
\cite{King:2009mk} are also considered for neutrino mixing besides
the above mentioned simple groups, the extension of the discrete flavor symmetry to
the quark sector and grand unification theory (GUT) have been
investigated as well \cite{Altarelli:2008bg,Morisi:2007ft}, please see
Refs.\cite{Altarelli:2010gt,Ishimori:2010au} for a review. In this
work, we shall study another 39 element simple discrete group
$T_{13}$ in flavor model building, which has gained much less
attention.

Recently 76 discrete groups with 3-dimensional representation were
scanned, it is suggested that $T_{13}$ is the group with the largest
fraction of TB mixing models \cite{Parattu:2010cy}. But the authors
set all the couplings to be equal to 1, the vacuum expectation
values are chosen to be 0 or 1, and the vacuum alignment is not
considered dynamically in Ref.\cite{Parattu:2010cy}. It is very
interesting to investigate the possible consistent realizations of
TB mixing based on $T_{13}$ group from this point of view. As far as
we know, the $T_{13}$ group as a discrete flavor symmetry has not
been discussed extensively. We note that a $T_{13}$ flavor model was
put forward in Ref. \cite{Kajiyama:2010sb}, and its implication in
the indirect detection of dark matter was studied. However, the
motivation is not to produce the TB mixing \footnote{The vacuum
alignment and the next leading order correction are not discussed in
Ref.\cite{Kajiyama:2010sb}, a set of numerical values are chosen by
hand for the model parameters so that the resulting lepton masses
and flavor mixing are consistent with experimental data.}. We have
tried many possible assignments for the involved fields, we find the
TB mixing can be produced exactly at leading order (LO) in some
scenarios, but meanwhile we face the difficulties that the first and
third light neutrino are degenerate or the corresponding vacuum
alignment is very difficult to be realized or some other problems.
In particularly, the realizations of TB mixing based on $T_{13}$
symmetry are drastically constrained after taking into account the
vacuum alignment issue. After lots of trial and error, we
construct a $T_{13}$ flavor model described in this work, where TB
mixing is obtained exactly at LO.  It is well-known that discrete
group $Z_N$ or continuous one like $U(1)$ are usually introduced to
eliminate unwanted couplings, to ensure the need vacuum alignment
and to reproduce the observed charged charged lepton mass
hierarchies. In the present work, the auxiliary symmetry $Z_4\times
Z_2$ is introduced for this purpose. It is notable that the charged
lepton mass hierarchies are determined by the $T_{13}\times
Z_4\times Z_2$ flavor symmetry itself without invoking a
Froggatt-Nielsen $U(1)$ symmetry.

This paper is organized as follows. In section 2, we discuss the
relevant features of $T_{13}$ group. In section 3, the structure of
the model is described, the LO results for neutrino as well as
charged lepton mass matrices are presented. In section 4, we show
how to get in a natural way the required vacuum alignment used
throughout the paper. In section 5, we present the study on the
corrections introduced by the higher order terms, which is
responsible for the deviation from TB mixing. Finally
section 6 is devoted to our conclusion. We give the explicit
representation matrices and the Clebsch-Gordan coefficients of
$T_{13}$ group in Appendix A. The analysis of the subleading
corrections to the vacuum alignment is presented in Appendix B.

\section{The discrete group $T_{13}$}

The discrete group $T_{13}$ is a subgroup of $SU(3)$, and it is
smallest discrete group with two complex irreducible
three-dimensional representations. $T_{13}$ is isomorphic to
$Z_{13}\rtimes Z_3$ \cite{Fairbairn:1982jx,book}, consequently it
has 39 group elements. $T_{13}$ can be generated by two elements $S$
and $T$ obeying the relations
\begin{equation}
\label{g1}S^{13}=T^3=1,~~~ST=TS^3
\end{equation}
The 39 elements of the group belong to 7 conjugate classes and are
generated from $S$ and $T$ as follows,
\begin{eqnarray}
\nonumber&& {\cal C}_1: e\\
\nonumber&& {\cal C}_2:\,T, TS, TS^2, TS^3, TS^4, TS^5, TS^6, TS^7,
TS^8, TS^9, TS^{10}, TS^{11}, TS^{12}\\
\nonumber&& {\cal C}_3:\,T^2, T^2S, T^2S^2, T^2S^3, T^2S^4, T^2S^5,
T^2S^6, T^2S^7, T^2S^8, T^2S^9, T^2S^{10}, T^2S^{11}, T^2S^{12}\\
\nonumber&& {\cal C}_4: S, S^3, S^9\\
\nonumber&& {\cal C}_5: S^4, S^{10}, S^{12}\\
\nonumber&& {\cal C}_6: S^2, S^5, S^6\\
\label{g2}&& {\cal C}_7: S^7, S^8, S^{11}
\end{eqnarray}
The $T_{13}$ group has 7 inequivalent irreducible representations
$\mathbf{1_1}$, $\mathbf{1_2}$, $\mathbf{1_3}$, $\mathbf{3_1}$,
$\mathbf{\bar{3}_1}$, $\mathbf{3_2}$ and $\mathbf{\bar{3}_2}$. It is
easy to see that the one-dimensional representations are given by
\begin{eqnarray}
\nonumber&&\mathbf{1_1}:~~S=1,~~T=1\\
\nonumber&&\mathbf{1_2}:~~S=1,~~T=\omega\\
\label{1}&&\mathbf{1_3}:~~S=1,~~T=\omega^2
\end{eqnarray}
where $\omega=e^{2i\pi/3}$. The three-dimensional representations
are given by
\begin{eqnarray}
\nonumber&&\mathbf{3_1}:~~S=\left(\begin{array}{ccc}\rho&0
&0\\
0&\rho^3&0 \\
0&0&\rho^9
\end{array}\right),~~~T=\left(\begin{array}{ccc}0&0&1\\
1&0&0\\
0&1&0
\end{array}\right)\\
\label{2}&&\mathbf{3_2}:~~S=\left(\begin{array}{ccc}\rho^2&0&0\\
0&\rho^6&0\\
0&0&\rho^5
\end{array}\right),~~T=\left(\begin{array}{ccc}0&0&1\\
1&0&0\\
0&1&0
\end{array}\right)
\end{eqnarray}
where $\rho=e^{2i\pi/13}$, the $\mathbf{\bar{3}_1}$ and
$\mathbf{\bar{3}_2}$ representations can be obtained by performing
the complex conjugation of $\mathbf{3_1}$ and $\mathbf{3_2}$
respectively. We can straightforwardly calculate the character table
of $T_{13}$, which is shown in Table \ref{tab:character}. Then the
multiplication rules between various representations follow
immediately,
\begin{table}[t]
\begin{center}
\begin{tabular}{|c|c|c|c|c|c|c|c|}\hline\hline
   &\multicolumn{7}{|c|}{classes}\\\cline{2-8}
   &${\cal C}_1$&${\cal C}_2$&${\cal C}_3$&${\cal C}_4$&${\cal
C}_5$ &${\cal C}_6$ &${\cal C}_7$\\\hline

$n_{{\cal C}_i}$&1&13&13&3&3&3&3\\\hline

$h_{{\cal C}_i}$&1&3&3&13&13&13&13\\\hline

$\mathbf{1_1}$&1&1&1&1&1&1&1\\\hline

$\mathbf{1_2}$&1&$\omega$ & $\omega^2$&1&1&1&1\\\hline

$\mathbf{1_3}$&1&$\omega^2$ & $\omega$&1&1&1&1\\\hline

$\mathbf{3_1}$&3&0&0&$\xi_1$& $\xi^{*}_1$ & $\xi_2$& $\xi^{*}_2$
\\\hline

$\mathbf{\bar{3}_1}$&3&0&0&$\xi^{*}_1$& $\xi_1$ & $\xi^{*}_2$&
$\xi_2$
\\\hline

$\mathbf{3_2}$&3&0&0&$\xi_2$& $\xi^{*}_2$ & $\xi^{*}_1$ &
$\xi_1$\\\hline

$\mathbf{\bar{3}_2}$&3&0&0&$\xi^{*}_2$& $\xi_2$ & $\xi_1$ &
$\xi^{*}_1$\\\hline\hline

\end{tabular}
\caption{\label{tab:character}Character table of the $T_{13}$ group,
where $\xi_1=\rho+\rho^3+\rho^9$, $\xi_2=\rho^2+\rho^5+\rho^6$,
$\rho=e^{2i\pi/13}$ and $\omega=e^{2i\pi/3}$. $n_{{\cal C}_i}$
denotes the number of the elements contained in the class ${\cal
C}_i$, and $h_{{\cal C}_i}$ is the order of the elements of ${\cal
C}_i$.}
\end{center}
\end{table}
\begin{eqnarray}
\nonumber&&\mathbf{1_1}\otimes
R=R\otimes\mathbf{1_1}=R,~~\mathbf{1_2}\otimes\mathbf{1_2}=\mathbf{1_3},~~\mathbf{1_2}\otimes\mathbf{1_3}=\mathbf{1_1},~~\mathbf{1_3}\otimes\mathbf{1_3}=\mathbf{1_2},\\
\nonumber&&\mathbf{1}_i\otimes\mathbf{3_1}=\mathbf{3_1},~~\mathbf{1}_i\otimes\mathbf{\bar{3}_1}=\mathbf{\bar{3}_1},~~\mathbf{1}_i\otimes\mathbf{3_2}=\mathbf{3_2},~~\mathbf{1}_i\otimes\mathbf{\bar{3}_2}=\mathbf{\bar{3}_2},\\
\nonumber&&\mathbf{3_1}\otimes\mathbf{3_1}=\mathbf{\bar{3}}_{1S}\oplus\mathbf{\bar{3}}_{1A}\oplus\mathbf{3_2},~~\mathbf{3_1}\otimes\mathbf{\bar{3}_1}=\mathbf{1_1}\oplus\mathbf{1_2}\oplus\mathbf{1_3}\oplus\mathbf{3_2}\oplus\mathbf{\bar{3}_2},\\
\nonumber&&\mathbf{3_1}\otimes\mathbf{3_2}=\mathbf{3_1}\oplus\mathbf{3_2}\oplus\mathbf{\bar{3}_2},~~\mathbf{3_1}\otimes\mathbf{\bar{3}_2}=\mathbf{3_1}\oplus\mathbf{\bar{3}_1}\oplus\mathbf{\bar{3}_2},\\
\nonumber&&\mathbf{\bar{3}_1}\otimes\mathbf{\bar{3}_1}=\mathbf{3}_{1S}\oplus\mathbf{3}_{1A}\oplus\mathbf{\bar{3}_2},~~\mathbf{\bar{3}_1}\otimes\mathbf{3_2}=\mathbf{3_1}\oplus\mathbf{\bar{3}_1}\oplus\mathbf{3_2},\\
\nonumber&&\mathbf{\bar{3}_1}\otimes\mathbf{\bar{3}_2}=\mathbf{\bar{3}_1}\oplus\mathbf{3_2}\oplus\mathbf{\bar{3}_2},~~\mathbf{3_2}\otimes\mathbf{3_2}=\mathbf{\bar{3}_1}\oplus\mathbf{\bar{3}}_{2S}\oplus\mathbf{\bar{3}}_{2A},\\
\label{g3}&&\mathbf{3_2}\otimes\mathbf{\bar{3}_2}=\mathbf{1_1}\oplus\mathbf{1_2}\oplus\mathbf{1_3}\oplus\mathbf{3_1}\oplus\mathbf{\bar{3}_1},~~\mathbf{\bar{3}_2}\otimes\mathbf{\bar{3}_2}=\mathbf{3_1}\oplus\mathbf{3}_{2S}\oplus\mathbf{3}_{2A}
\end{eqnarray}
where the indices $i=2,\,3$, $R$ indicates any $T_{13}$ irreducible
representation, and the subscript $S$ and $A$ denote symmetric and
anti-symmetric products respectively. The explicit representation
matrices of the group elements for the three dimensional irreducible
representations are listed in Appendix A. From these representation
matrices, one can directly calculate the Clebsch-Gordan coefficients
for the decomposition of the product representations, which are
given in Appendix A as well.

\section{The structure of the model}

The model is supersymmetric and based on the discrete symmetry
$T_{13}\times Z_4\times Z_2$. Supersymmetry (SUSY) is introduced in
order to simplify the discussion of the vacuum alignment. All the
fields of the model, together with their transformation properties
under the flavor group, are listed in Table \ref{tab:field}. We
assign the 3 generation of left-handed lepton doublets $\ell$ to be
the $\mathbf{3_1}$ representation, while the right-handed charged
lepton $e^c$, $\mu^c$ and $\tau^c$ transform as $\mathbf{1_1}$,
$\mathbf{1_2}$ and $\mathbf{1_3}$ respectively. It is notable that
the three right-handed neutrinos $\nu^{c}_1$, $\nu^{c}_2$ and
$\nu^{c}_3$ are assigned as $\mathbf{1_1}$, $\mathbf{1_2}$ and
$\mathbf{1_3}$ as well, they transform in the same way as the
right-handed charged lepton fields. This is an interesting feature
of the model. We note that in popular $A_4$ and $S_4$ models, the
right-handed neutrinos are frequently treated to be a triplet
\cite{TBModel,S4model}. Lepton masses and mixing arise from the
spontaneous breaking of the flavor symmetry by means of the flavon
fields, they are neutral under the standard model gauge group and
are divided into two sets $\Phi_{\ell}=\{\chi, \xi\}$ and
$\Phi_{\nu}=\{\phi, \eta\}$. We note that all the flavon fields are
triplets under $T_{13}$ in this work, $\Phi_{\ell}$ is responsible
for the charged lepton masses and $\Phi_{\nu}$ for the neutrino
masses at LO. In the following, we shall discuss the LO predictions
for fermion masses and flavor mixings. For the time being we assume
that the scalar components of the flavon fields acquire vacuum
expectation values (VEV) according to the following scheme,
\begin{eqnarray}
\nonumber&&
\langle\chi\rangle=v_{\chi}\left(\begin{array}{c}1\\
1\\
1
\end{array}\right),~~~~~~~~\langle\xi\rangle=v_{\xi}\left(\begin{array}{c}1\\
1\\
1\end{array}\right)\\
\label{3}&&\langle\phi\rangle=v_{\phi}\left(\begin{array}{c}0\\
1\\
-1
\end{array}\right),~~~~~~~\langle\eta\rangle=\left(\begin{array}{c}0\\
v_{\eta}\\
0\end{array}\right)
\end{eqnarray}
\begin{table}[t]
\begin{center}
\begin{tabular}{|c|c|c|c|c|c|c|c|c||c|c|c|c||c|c|c|c|c|}\hline\hline
Fields & $\ell$ & $e^c$ & $\mu^c$  &  $\tau^c$ & $\nu^{c}_1$ &
$\nu^{c}_2$  &  $\nu^{c}_3$ &$h_{u,d}$ & $\chi$ & $\xi$ & $\phi$  &
$\eta$ & $\chi^{0}$ & $\rho^0$ & $\theta^{0}$ & $\eta^0$ &
$\xi^{0}$\\\hline

$T_{13}$ & $\mathbf{3_1}$ & $\mathbf{1_1}$ & $\mathbf{1_2}$ &
$\mathbf{1_3}$ & $\mathbf{1_1}$  & $\mathbf{1_2}$
 & $\mathbf{1_3}$ & $\mathbf{1_1}$ & $\mathbf{\bar{3}_1}$ & $\mathbf{3_1}$ & $\mathbf{\bar{3}_1}$ &  $\mathbf{\bar{3}_2}$ & $\mathbf{3_2}$ & $\mathbf{1_2}$ & $\mathbf{1_3}$ &$\mathbf{\bar{3}_2}$ & $\mathbf{1_1}$\\\hline

$Z_4$  &  1  & i & -1  & -i  & 1 & 1 &1 & 1 & i & i  &  1  & 1 &-1 &
-1& -1 & 1 & -i  \\\hline

$Z_2$ &  1  &1  & 1  &1  & -1 & -1 & -1  & 1 & 1 & 1 & -1 & -1 & 1 &
1 & 1 &1 &-1\\ \hline

$U(1)_R$ & 1 & 1 & 1& 1 & 1 & 1 & 1 & 0& 0 & 0 & 0 & 0 &2 & 2 & 2 & 2 & 2 \\ \hline\hline
\end{tabular}
\caption{\label{tab:field}The transformation properties of the
matter fields, the electroweak Higgs doublets, the flavon fields and
the driving fields under the flavor symmetry $T_{13}\times Z_4\times
Z_2$.}
\end{center}
\end{table}
In section 4 we shall show that the above alignment is indeed
naturally realized at LO from the most general superpotential
allowed by the symmetry of the model.

\subsection{Charged leptons}
The charged lepton masses are described by the following
superpotential
\begin{eqnarray}
\label{4}w_{\ell}=\sum^{5}_{i=1}\frac{y_{e_i}}{\Lambda^3}e^{c}(\ell{\cal
O}_{i})_{\mathbf{1_1}}h_d
+\frac{y'_{\mu}}{\Lambda^2}\mu^{c}(\ell(\xi\xi)_{\mathbf{\bar{3}}_{1S}})_{\mathbf{1_3}}h_d
+\frac{y_{\tau}}{\Lambda}\tau^{c}(\ell\chi)_{\mathbf{1_2}}h_d+...
\end{eqnarray}
where
\begin{equation}
\label{5}{\cal
O}=\{(\chi(\chi\chi)_{\mathbf{\bar{3}_{2}}})_{\mathbf{\bar{3}_1}},\,
((\chi\chi)_{\mathbf{3}_{1S}}\xi)_{\mathbf{\bar{3}}_{1S}},\,((\chi\chi)_{\mathbf{3}_{1S}}\xi)_{\mathbf{\bar{3}}_{1A}},\,((\chi\chi)_{\mathbf{\bar{3}_2}}\xi)_{\mathbf{\bar{3}_1}},\,(\chi(\xi\xi)_{\mathbf{3_2}})_{\mathbf{\bar{3}_1}}\}
\end{equation}
We note that the subscripts $\mathbf{1_1}$, $\mathbf{1_2}$,
$\mathbf{1_3}$, $\mathbf{\bar{3}_1}$ etc denote the $T_{13}$
contractions. In the above superpotential $w_{\ell}$, for each
charged lepton, only the lowest order operators in the expansion in
powers of $1/\Lambda$ are displayed explicitly. Dots stand for
higher dimensional operators which will be discussed later. It is
remarkable that the $Z_4$ symmetry imposes different powers of
$\chi$ and $\xi$ for the electron, muon and tauon terms, i.e., only
the tau mass is generated at LO, the muon and the electron masses
are generated by high order contributions. After electroweak and
flavor symmetry breaking, we have
\begin{eqnarray}
\nonumber
&&w_{\ell}=[y_{e_1}\frac{v^3_{\chi}}{\Lambda^3}+4y_{e_2}\frac{v^2_{\chi}v_{\xi}}{\Lambda^3}+y_{e_4}\frac{v^2_{\chi}v_{\xi}}{\Lambda^3}+y_{e_5}\frac{v_{\chi}v^2_{\xi}}{\Lambda^3}]
v_de^c(e+\mu+\tau)+2y'_{\mu}\frac{v^2_{\xi}}{\Lambda^2}v_{d}\mu^c(e+\omega^2\mu+\omega\tau)\\
\nonumber&&~~~+y_{\tau}\frac{v_{\chi}}{\Lambda}v_d\tau^c(e+\omega\mu+\omega^2\tau)\\
\label{6}&&~~\equiv y_e\frac{v^3_{\chi}}{\Lambda^3}
v_de^c(e+\mu+\tau)+y_{\mu}\frac{v^2_{\xi}}{\Lambda^2}v_d\mu^c(e+\omega^2\mu+\omega\tau)+y_{\tau}\frac{v_{\chi}}{\Lambda}v_d\tau^c(e+\omega\mu+\omega^2\tau)
\end{eqnarray}
where $v_d=\langle h_d\rangle$, the parameters $y_e$ and $y_{\mu}$
are parameterized as
$y_e=y_{e_1}+(4y_{e_2}+y_{e_4})v_{\xi}/v_{\chi}+y_{e_5}v^2_{\xi}/v^2_{\chi}$
and $y_{\mu}=2y'_{\mu}$. As a result, the charged lepton mass matrix
has the form
\begin{eqnarray}
\nonumber m_{\ell}&=&\left(\begin{array}{ccc}y_e\frac{v^3_{\chi}}{\Lambda^3}&y_e\frac{v^3_{\chi}}{\Lambda^3}&y_e\frac{v^3_{\chi}}{\Lambda^3}\\
y_{\mu}\frac{v^2_{\xi}}{\Lambda^2} &
\omega^2y_{\mu}\frac{v^2_{\xi}}{\Lambda^2} & \omega
y_{\mu}\frac{v^2_{\xi}}{\Lambda^2}\\
y_{\tau}\frac{v_{\chi}}{\Lambda} & \omega
y_{\tau}\frac{v_{\chi}}{\Lambda} & \omega^2
y_{\tau}\frac{v_{\chi}}{\Lambda}
\end{array}\right)v_d\\
\label{8}&=&\left(\begin{array}{ccc}y_e\frac{v^3_{\chi}}{\Lambda^3}&0
&0 \\
0& y_{\mu}\frac{v^2_{\xi}}{\Lambda^2} &0 \\
0&0& y_{\tau}\frac{v_{\chi}}{\Lambda}
\end{array}\right)\left(\begin{array}{ccc}1&1&1\\
1&\omega^2&\omega\\
1& \omega & \omega^2
\end{array}\right)v_d
\end{eqnarray}
Obviously the charged lepton mass matrix is diagonalized by
performing the transformation $\ell\rightarrow U_{\ell}\ell$, where
$U_{\ell}$ is
\begin{equation}
\label{9}U_{\ell}=\frac{1}{\sqrt{3}}\left(\begin{array}{ccc}1&1&1\\
1&\omega&\omega^2\\
1&\omega^2& \omega
\end{array}\right)
\end{equation}
and the charged lepton masses are given by
\begin{equation}
\label{10}m_e=\sqrt{3}\,\Big|y_e\frac{v^3_{\chi}}{\Lambda^3}v_d\Big|,~~~m_{\mu}=\sqrt{3}\,\Big|y_{\mu}\frac{v^2_{\xi}}{\Lambda^2}v_d\Big|,~~~m_{\tau}=\sqrt{3}\,\Big|y_{\tau}\frac{v_{\chi}}{\Lambda}v_d\Big|
\end{equation}
We see that the charged lepton mass hierarchies are generate by the
spontaneous breaking of the flavor symmetry. To estimate the order
of magnitudes of $v_{\chi}$ and $v_{\xi}$, we can use the
experimental data on the ratios of charged lepton masses. Assuming
that the coefficients $y_e$, $y_{\mu}$ and $y_{\tau}$ are of ${\cal
O}(1)$, we have
\begin{eqnarray}
\nonumber&&\frac{m_e}{m_{\tau}}\sim\frac{v^2_{\chi}}{\Lambda^2}\simeq
0.0003\\
\label{11}&&\frac{m_{\mu}}{m_{\tau}}\sim\frac{v^2_{\xi}}{v_{\chi}\Lambda}\simeq0.06
\end{eqnarray}
These relations are satisfied for
\begin{equation}
\label{12}(\frac{v_{\chi}}{\Lambda},\frac{v_{\xi}}{\Lambda})\sim(0.017,\pm0.032)
\end{equation}
we see that the amplitudes of both $v_{\chi}/\Lambda$ and
$v_{\xi}/\Lambda$ are roughly of the same order about $\lambda^2_c$,
where $\lambda_c$ is the Cabibbo angle. It is interesting to
investigate the flavor symmetry breaking pattern in the charged
lepton sector, it is induced by the VEVs of $\chi$ and $\xi$ at LO.
Given the explicit representation matrices listed in Appendix A, it
is obvious that the VEVs of $\chi$ and $\xi$ are invariant under the
action of $T$ and $T^2$. Furthermore, we can check that the
hermitian matrix $m^{\dagger}_{\ell}m_{\ell}$ is invariant under
both $T$ and $T^2$ as well. Therefore we conclude that the $T_{13}$ flavor
symmetry is broken down to the $Z_3$ subgroup generated by the element
$T$ in the charged lepton sector at LO.

\subsection{Neutrinos}
The superpotential for the neutrino sector can be written as
\begin{eqnarray}
\nonumber&&w^{SS}_{\nu}=\frac{y_{\nu1}}{\Lambda}\nu^{c}_1(\ell\phi)_{\mathbf{1_1}}h_u+\frac{y_{\nu2}}{\Lambda}\nu^{c}_2(\ell\phi)_{\mathbf{1_3}}h_u
+\frac{y_{\nu3}}{\Lambda}\nu^{c}_3(\ell\phi)_{\mathbf{1_2}}h_u+\frac{1}{2}M_1\nu^{c}_1\nu^{c}_1+\frac{1}{2}M_2(\nu^{c}_2\nu^{c}_3+\nu^{c}_3\nu^{c}_2)+...\\
\nonumber&&w^{eff}_{\nu}=\frac{x_{\nu1}}{\Lambda^3}((\ell h_u\ell
h_u)_{\mathbf{\bar{3}}_{1S}}(\phi\phi)_{\mathbf{3}_{1S}})_{\mathbf{1_1}}+\frac{x_{\nu2}}{\Lambda^3}((\ell
h_u\ell
h_u)_{\mathbf{\bar{3}}_{1S}}(\eta\eta)_{\mathbf{3_1}})_{\mathbf{1_1}}+\frac{x_{\nu3}}{\Lambda^3}((\ell
h_u\ell
h_u)_{\mathbf{3_2}}(\phi\phi)_{\mathbf{\bar{3}_2}})_{\mathbf{1_1}}\\
\label{13}&&~~+\frac{x_{\nu4}}{\Lambda^3}((\ell h_u\ell
h_u)_{\mathbf{3_2}}(\phi\eta)_{\mathbf{\bar{3}_2}})_{\mathbf{1_1}}+...
\end{eqnarray}
where $M_1$ and $M_2$ are constants with dimension of mass, they are
naturally of the same order as the cutoff scale $\Lambda$, and the
factor of $\frac{1}{2}$ is a normalization factor for convenience.
We note that $w^{SS}_{\nu}$ denotes the lagrangian of the type I
see-saw mechanism, and $w^{eff}_{\nu}$ is the collection of higher
dimensional Weinberg operators. Taking into account the vacuum
alignment shown in Eq.(\ref{3}), we can read the Dirac and Majorana
mass matrices immediately from $w^{SS}_{\nu}$ as follows
\begin{equation}
\label{14}m_D=\left(\begin{array}{ccc}0&y_{\nu1}&-y_{\nu1}\\
0&\omega^2y_{\nu2}& -\omega y_{\nu2}\\
0& \omega y_{\nu3}& -\omega^2y_{\nu3}
\end{array}\right)\frac{v_{\phi}}{\Lambda}v_{u},~~~~m_{M}=\left(\begin{array}{ccc}M_1&0&0\\
0&0& M_2\\
0& M_2&0
\end{array}\right)
\end{equation}
where $v_{u}$ is the vacuum expectation value of the Higgs field
$h_u$. It is remarkable that the eigenvalues of the Majorana mass
matrix $m_M$ are $M_1$, $M_2$ and $-M_2$, two of the right handed
neutrinos are degenerate at LO. This is a distinguished feature of
our model from the previous flavor models in which the right-handed
neutrinos are usually treated to form a triplet. It is very
interesting to discuss the assignment of right-handed neutrinos as
singlets and the corresponding phenomenological implications in
flavor models based on $A_4$, $\Delta(27)$, $S_4$ and so on. The
light neutrino mass matrix from see-saw mechanism is given by the
well-known see-saw formula
\begin{equation}
\label{15}m^{SS}_{\nu}=-m^T_Dm^{-1}_Mm_D=\left(\begin{array}{ccc}0&0&0\\
0&-a-2b &a-b\\
0&a-b&-a-2b
\end{array}\right)\frac{v^2_u}{\Lambda}  
\end{equation}
where
\begin{equation}
\label{16}a=y^2_{\nu1}v^2_{\phi}/(\Lambda M_1),~~~~~~b=y_{\nu2}y_{\nu3}v^2_{\phi}/(\Lambda M_2) 
\end{equation}
The superpotential $w^{eff}_{\nu}$ leads to the following effective
light neutrino mass matrix
\begin{equation}
\label{17}m^{eff}_{\nu}=\left(\begin{array}{ccc} r &0&0\\
0 &s &t\\
0&t&s
\end{array}\right)\frac{v^2_{u}}{\Lambda}
\end{equation}
where
\begin{eqnarray}
\nonumber&&r=-2x_{\nu4}v_{\eta}v_{\phi}/\Lambda^2\\
\nonumber&&s=2x_{\nu3}v^2_{\phi}/\Lambda^2\\
\label{18}&&t=-4x_{\nu1}v^2_{\phi}/\Lambda^2+2x_{\nu2}v^2_{\eta}/\Lambda^2
\end{eqnarray}
Therefore in the flavor basis where the charged lepton mass matrix
is diagonal, the light neutrino mass matrices read
\begin{eqnarray}
\nonumber&&m^{SS}_{\nu}=\left(\begin{array}{ccc} -2b &b& b\\
b&a & -a-b\\
b&-a-b &a
\end{array}\right)\frac{v^2_u}{\Lambda}\\
\label{19}&&m^{eff}_{\nu}=\left(\begin{array}{ccc}r+2s+2t&r-s-t& r-s-t\\
r-s-t & r-s+2t & r+2s-t\\
r-s-t & r+2s-t &r-s+2t
\end{array}\right)\frac{v^2_u}{3\Lambda}
\end{eqnarray}
Both the light neutrino mass matrices $m^{SS}_{\nu}$ and
$m^{eff}_{\nu}$ are $2\leftrightarrow3$ invariant, and they satisfy
the magic symmetry
$(m^{SS(eff)}_{\nu})_{11}+(m^{SS(eff)}_{\nu})_{13}=(m^{SS(eff)}_{\nu})_{22}+(m^{SS(eff)}_{\nu})_{32}$.
Therefore they are exactly diagonalized by the TB mixing matrix
\begin{eqnarray}
\nonumber&&U_{TB}m^{SS}_{\nu}U_{TB}={\rm
diag}(-3b,0,2a+b)\frac{v^2_u}{\Lambda}\\
\label{20}&&U_{TB}m^{eff}_{\nu}U_{TB}={\rm
diag}(s+t,r,-s+t)\frac{v^2_u}{\Lambda}
\end{eqnarray}
We note that the contribution $m^{SS}_{\nu}$ from the see-saw
mechanism is of the same order as $m^{eff}_{\nu}$ coming from the
Weinberg operators, consequently both contributions should be
included. The light neutrino mass matrix is the sum of
$m^{SS}_{\nu}$ and $m^{eff}_{\nu}$
\begin{equation}
\label{21}m_{\nu}=m^{SS}_{\nu}+m^{eff}_{\nu}
\end{equation}
Obviously $m_{\nu}$ is still diagonalized by the TB mixing matrix,
and the light neutrino masses are given by
\begin{eqnarray}
\nonumber&&
m_{\nu1}=(s+t-3b)\frac{v^2_u}{\Lambda}\\
\nonumber&&m_{\nu2}=r\frac{v^2_u}{\Lambda}\\
\label{22}&&m_{\nu3}=(-s+t+2a+b)\frac{v^2_u}{\Lambda}
\end{eqnarray}
where $U_{TB}$ is the well-known TB mixing matrix
\begin{equation}
\label{23}U_{TB}=\left(\begin{array}{ccc}\sqrt{\frac{2}{3}}
&\frac{1}{\sqrt{3}}  &0\\
-\frac{1}{\sqrt{6}}  & \frac{1}{\sqrt{3}}  & \frac{1}{\sqrt{2}}\\
-\frac{1}{\sqrt{6}}  &  \frac{1}{\sqrt{3}} &-\frac{1}{\sqrt{2}}
\end{array}\right)
\end{equation}
We note that the contributions proportional to $a$ and $b$ can be
absorbed into $s$ and $t$ by redefinition $s\rightarrow s-a-2b$ and
$t\rightarrow t+a-b$, therefore the light neutrino masses depend on
three unrelated complex parameters. There are more freedoms to tune
the mass differences and then satisfy the constraints associated to
neutrino oscillation, the neutrino mass spectrum can be normal
hierarchy or inverted hierarchy. In contrast with some "constrained"
flavor models, no neutrino mass sum-rules \cite{Barry:2010yk} can be
found in this model. We could certainly remove the right-handed
neutrinos from our model, then the neutrino masses are described by
the Weinberg operators $w^{eff}_{\nu}$, the above conclusions remain
invariant. However, if we only concentrate on the see-saw
realization $w^{SS}_{\nu}$, the second neutrino would be massless
although the lepton mixing is of TB form, this scenario is ruled out
by the experimental observations.

It is notable that the VEVs of $\phi$ and $\eta$ are always changed
under the action of any $T_{13}$ group element except unit element,
consequently the flavor symmetry $T_{13}$ is broken down to nothing
in the neutrino sector at LO. Reminding that ones usually break the
flavor symmetry into the low energy neutrino symmetry group Klein
four \cite{Bazzocchi:2009pv,Ding:2009iy,Ding:2010pc} or $Z_2$
\cite{Altarelli:2005yp,Altarelli:2005yx} to guarantee TB mixing for
neutrinos, it is really amazing we can still obtain TB mixing even
if the flavor symmetry is broken completely in the neutrino sector
at LO.

In short summary, at the LO the $T_{13}$ flavor symmetry is broken
down to $Z_3$ subgroup and nothing in the charged lepton and
neutrino sectors respectively. This breaking chain lets us to find
the TB scheme at LO as the lepton mixing matrix. However, the mixing
angles generally deviate from the TB values after the corrections of
the higher order terms are included. It is remarkable that this
symmetry breaking pattern of our model has not been studied, as far
as we know. It is attractive to investigate whether we can still
reproduce TB mixing in models with $A_4$ or $S_4$ symmetry, if the
flavor symmetry is broken completely in the neutrino sector at LO.

\section{Vacuum alignment}
The vacuum alignment problem of the model can be solved by the
supersymmetric driving fields method introduced in
Ref.\cite{Altarelli:2005yx}. This approach exploits a continuous
$U(1)_R$ symmetry under which matter fields have $R=+1$, while
Higgses and flavon fields have $R=0$. Such a symmetry will be
eventually broken down to the R-parity by small SUSY breaking
effects which can be neglected in the first approximation in our
analysis. The spontaneous breaking of $T_{13}$ can be employed by
introducing the so-called driving fields with $R=2$, which enter
linearly into the superpotential. Five driving fields $\chi^{0}$,
$\rho^0$, $\theta^{0}$, $\eta^0$ and $\xi^0$ are introduced in our
model, their transformation rules under the flavor symmetry are
shown in Table \ref{tab:field}. We note that the driving fields
$\rho^{0}$ and $\theta^{0}$ are necessary to stabilize the vacuum
alignment under subleading corrections. At LO, the most general
superpotential dependent on the driving fields, which is invariant
under the flavor symmetry group $T_{13}\times Z_4\times Z_2$, is
given by
\begin{eqnarray}
\nonumber&&
w_v=f_1(\chi^0(\chi\chi)_{\mathbf{\bar{3}_2}})_{\mathbf{1_1}}+f_2(\chi^0(\chi\xi)_{\mathbf{\bar{3}_2}})_{\mathbf{1_1}}+f_3\rho^{0}(\chi\xi)_{\mathbf{1_3}}+f_4\theta^{0}(\chi\xi)_{\mathbf{1_2}}
+g_1(\eta^{0}(\eta\eta)_{\mathbf{3}_{2S}})_{\mathbf{1_1}}\\
\label{24}&&~~~+g_2(\eta^0(\phi\eta)_{\mathbf{3_2}})_{\mathbf{1_1}}+h\xi^{0}(\xi\phi)_{\mathbf{1_1}}
\end{eqnarray}
In the SUSY limit, the vacuum configuration is determined by the
vanishing of the derivative of $w_v$ with respect to each component
of the driving fields
\begin{subequations}
\begin{eqnarray}
\label{25a}&& \frac{\partial
w_v}{\partial\chi^0_1}=f_1\chi^2_1+f_2\chi_2\xi_1=0\\
\label{25b}&&\frac{\partial
w_v}{\partial\chi^0_2}=f_1\chi^2_2+f_2\chi_3\xi_2=0\\
\label{25c}&&\frac{\partial
w_v}{\partial\chi^0_3}=f_1\chi^2_3+f_2\chi_1\xi_3=0
\end{eqnarray}
\end{subequations}\vskip-0.3in
\begin{eqnarray}
\label{add2}&&~~~~~~~~~~~~~~~~~\frac{\partial
w_v}{\partial\rho^{0}}=f_3(\chi_1\xi_1+\omega^2\chi_2\xi_2+\omega\chi_3\xi_3)=0\\
\label{add3}&&~~~~~~~~~~~~~~~~~\frac{\partial
w_v}{\partial\theta^{0}}=f_4(\chi_1\xi_1+\omega\chi_2\xi_2+\omega^2\chi_3\xi_3)=0
\end{eqnarray}
The above equations are satisfied by the alignment
\begin{equation}
\label{26}\langle\chi\rangle=v_{\chi}\left(\begin{array}{c}1\\
1\\
1\end{array}\right),~~~~~~~\langle\xi\rangle=v_{\xi}\left(\begin{array}{c}1\\
1\\
1\end{array}\right)
\end{equation}
with the relation
\begin{equation}
\label{27}v_{\chi}=-\frac{f_2}{f_1}v_{\xi},~~~~v_{\xi}~ {\rm
undetermined}
\end{equation}
Without assuming any fine-tuning among the parameters $f_1$ and
$f_2$, the VEVs $v_{\chi}$ and $v_{\xi}$ are expected to be of the
same order of magnitude, this is consistent with the conclusion drew
from the charged lepton mass hierarchies. We note that if one
component of $\chi$ or $\xi$ has vanishing VEV, e.g.
$\langle\xi_1\rangle=0$, Eqs.(\ref{25a})-(\ref{25c}) imply
$\langle\chi_1\rangle=\langle\chi_2\rangle=\langle\chi_3\rangle=0$.
This means that the VEV of any component of the flavons $\chi$ or
$\xi$ should be non-zero in order to obtain a non-trivial vacuum
configuration. As has been shown in the previous section, at LO the
$T_{13}$ flavor symmetry is spontaneously broken by the VEVs of
$\phi$ and $\eta$ in the neutrino sector, their vacuum
configurations are determined by
\begin{subequations}
\begin{eqnarray}
\label{28a}&&\frac{\partial
w_v}{\partial\eta^{0}_1}=2g_1\eta_2\eta_3+g_2\phi_3\eta_1=0\\
\label{28b}&&\frac{\partial
w_v}{\partial\eta^0_2}=2g_1\eta_1\eta_3+g_2\phi_1\eta_2=0\\
\label{28c}&&\frac{\partial
w_v}{\partial\eta^0_3}=2g_1\eta_1\eta_2+g_2\phi_2\eta_3=0
\end{eqnarray}
\end{subequations}\vskip-0.2in
\begin{equation}
\label{29}~~~~~~~~~~~~\frac{\partial
w_v}{\partial\xi^0}=h(\xi_1\phi_1+\xi_2\phi_2+\xi_3\phi_3)=0
\end{equation}
The first three equations Eq.(\ref{28a})-(\ref{28c}) lead to two
un-equivalent vacuum configurations \footnote{We note that the
equations can be satisfied by two additional solutions as well. One
is
\begin{equation}
\label{f1}\langle\phi\rangle=\left(\begin{array}{c}v_{\phi_1}\\0\\v_{\phi_3}\end{array}\right),~~~~~~~\langle\eta\rangle=\left(\begin{array}{c}0\\0\\v_{\eta}\end{array}\right)
\end{equation}
Another one is
\begin{equation}
\label{f2}\langle\phi\rangle=\left(\begin{array}{c}v_{\phi_1}\\
v_{\phi_2}\\0\end{array}\right),~~~~~~~\langle\eta\rangle=\left(\begin{array}{c}v_{\eta}\\0\\0\end{array}\right)
\end{equation}
where $v_{\eta}$, $v_{\phi_1}$, $v_{\phi_2}$ and $v_{\phi_3}$ are
undetermined. However, the above two solutions can be obtained by
acting on the vacuum Eq.(\ref{32}) with the elements $T$ and $T^2$
respectively. Therefore these two solutions are equivalent to the
configuration in Eq.(\ref{32}). }, the first is
\begin{equation}
\label{30}\langle\phi\rangle=v_{\phi}\left(\begin{array}{c}1\\
1\\
1\end{array}\right),~~~~~~~\langle\eta\rangle=v_{\eta}\left(\begin{array}{c}1\\
1\\
1\end{array}\right)
\end{equation}
with
\begin{equation}
\label{31}v_{\eta}=-\frac{g_2}{2g_1}v_{\phi},~~~~v_{\phi}~ {\rm
undetermined}
\end{equation}
The second solution is
\begin{equation}
\label{32}\langle\phi\rangle=\left(\begin{array}{c}0\\
v_{\phi_2}\\
v_{\phi_3}
\end{array}\right),~~~~~~~\langle\eta\rangle=\left(\begin{array}{c}0\\v_{\eta}\\0\end{array}\right)
\end{equation}
where $v_{\eta}$, $v_{\phi_2}$ and $v_{\phi_3}$ are constrained. 
Using the alignment of $\chi$ in Eq.(\ref{26}), for the first
solution shown in Eq.(\ref{30}), we can immediately infer from
Eq.(\ref{29})
\begin{equation}
\label{add1}v_{\xi}v_{\phi}=0
\end{equation}
We are led to the trivial solutions $v_{\chi}=v_{\xi}=0$ or
$v_{\phi}=v_{\eta}=0$, which can be removed by the interplay of
radiative corrections to the scalar potential and soft SUSY breaking
terms for the flavon fields. Therefore we choose the second solution
in this work, this vacuum configuration can produce the results in
the previous section. Then the minimization equation Eq.(\ref{29})
implies
\begin{equation}
\label{33}v_{\phi_2}+v_{\phi_3}=0
\end{equation}
This indicates that $\langle\phi_2\rangle$ and
$\langle\phi_3\rangle$ have to be equal up to a relative sign, thus
$\phi$ is fully aligned as
\begin{equation}
\label{34}\langle\phi\rangle=v_{\phi}\left(\begin{array}{c}0\\1\\
-1\end{array}\right)
\end{equation}
Starting from the vacuum configurations given in Eq.(\ref{26}),
Eq.(\ref{32}) and Eq.(\ref{34}) and acting on them with the elements
of the flavor symmetry group $T_{13}$, we can generate other minima
of the scalar potential. However, these new minima are physically
equivalent to the original one, it is not restrictive to analyze the
model by choosing the vacuum in Eqs.(\ref{26},\ref{32},\ref{34}) as
local minimum. It is important to check the stability of the LO vacuum configuration, if we introduce small perturbations to the VEVs of the flavon fields as follows,
\begin{eqnarray}
\nonumber&&\langle\chi\rangle=v_{\chi}\left(\begin{array}{c}1+x_1\\
1+x_2\\
1+x_3
\end{array}
\right),~~~~\langle\xi\rangle=v_{\xi}\left(\begin{array}{c}
1\\
1+y_2\\
1+y_3
\end{array}\right)\\
&&\langle\phi\rangle=v_{\phi}\left(\begin{array}{c}z_1\\
1\\
-1+z_3
\end{array}\right),~~~~\langle\eta\rangle=v_{\eta}\left(\begin{array}{c}
w_1\\
1\\
w_3
\end{array}\right)
\end{eqnarray}
After some straightforward algebra, we find that the only solution to the minimization equations is
\begin{eqnarray}
\nonumber&&x_1=x_2=x_3=0,~~~~y_2=y_3=0\\
&&z_1=z_3=0,~~~~w_1=w_3=0
\end{eqnarray}
Therefore the LO vacuum alignment is stable, then we turn to consider the magnitudes of flavon VEVs.
Since the VEVs of $\phi$ and $\eta$ are closely
related with each other through the equations
Eqs.(\ref{28a})-(\ref{28c}), and they have the same charges under
the auxiliary symmetry $Z_4\times Z_2$, we expect a common order of
magnitude for the VEVs $v_{\chi}$ and
$v_{\eta}$. 
However, the VEVs of $\Phi_{\ell}=\{\chi,\xi\}$ and
$\Phi_{\nu}=\{\phi,\eta\}$ can be in principle different and they
are subject to phenomenological constraints. As we have shown in
section 3.1, $\langle\Phi_{\ell}\rangle$ is responsible for the
charged lepton mass hierarchies, and it is required to satisfy
\begin{equation}
\label{35}\varepsilon\equiv\frac{v_{\chi}}{\Lambda}\sim\frac{v_{\xi}}{\Lambda}\sim\lambda^2_c
\end{equation}
Among the three neutrino mixing angles, the solar neutrino mixing
angle $\theta_{12}$ is measured most precisely so far, the
experimentally allowed departures of $\theta_{12}$ from its TB value
$\sin^2\theta_{12}=1/3$ are at most of order $\lambda^2_c$
\cite{Schwetz:2008er,Fogli:Indication,GonzalezGarcia:2010er}. It is
well-known that the superpotentials $w_{\ell}$, $w^{SS}_{\nu}$,
$w^{eff}_{\nu}$ and $w_v$ are corrected by higher dimensional
operators in the expansion (please see section 5 and Appendix B for
detail), which mostly can be constructed by including the
combination $\Phi_{\nu}\Phi_{\nu}$ on top of each LO term, thus all
the three mixing angles receive corrections of order
$\langle\Phi_{\nu}\rangle^2/\Lambda^2$ (please see section 5 for
detail). Requiring that the mixing angles particular $\theta_{12}$
lie in the ranges allowed by neutrino oscillation data, we obtain
the condition
\begin{equation}
\label{36}\varepsilon'\equiv\frac{v_{\phi}}{\Lambda}\sim\frac{v_{\eta}}{\Lambda}\leq\lambda_c
\end{equation}
The same condition follows from the requirement that the generated
charged lepton mass hierarchies should be stable under subleading
corrections. As a result, we can tolerate a moderate hierarchy
between $\varepsilon$ and $\varepsilon'$ because of the strong
constraint of the auxiliary symmetry $Z_4\times Z_2$. It is a
general conclusion that a hierarchy between the VEVs of the flavon
fields can be accommodated in a "fully" separated scalar potential.
This type of vacuum alignment is usually constructed to generate a
large reactor angle \cite{Lin:2009bw,Altarelli:2009gn }, i.e.
$\theta_{13}\sim\lambda_c$, although it is predicted to be exactly
zero at LO. However, the subleading corrections to $\theta_{13}$
turn out to be of order $\lambda^2_c$ in our model, as we shall
demonstrate in next section. 
\section{Subleading corrections}
It is crucial to guarantee that the successful LO predictions are
not spoiled by subleading corrections, we will discuss this
important issue in detail. The superpotentials $w_{\ell}$,
$w^{SS}_{\nu}$, $w^{eff}_{\nu}$ and $w_{v}$ are corrected by higher
dimensional operators, which arise from adding the products
$\Phi_{\nu}\Phi_{\nu}$, invariant combination under $Z_4\times Z_2$,
on top of the LO terms. Then the residual $Z_3$ symmetry in the
charged lepton sector would be broken completely by the subleading
contributions. The lepton masses and mixing matrices are corrected
by both the shift of the vacuum configuration and the higher
dimensional operators in the Yukawa superpotentials. As a result,
the mass matrices with subleading corrections can be obtained by
inserting the modified vacuum alignment into the LO Yukawa operators
plus the contributions of the new higher dimensional operators
evaluated with the unperturbed VEVs.

The subleading corrections to the vacuum alignment are discussed in
Appendix B in detail. The inclusion of the higher dimensional
operators in the driving superpotential $w_v$ results in a shift of
the VEVs of the flavon fields, the vacuum configuration is modified
into
\begin{eqnarray}
\nonumber&&\langle\chi\rangle=\left(\begin{array}{c} v_{\chi}+\delta
v_{\chi_1}\\
v_{\chi}+\delta
v_{\chi_2}\\
v_{\chi}+\delta v_{\chi_3}
\end{array}\right),~~~~~~~\langle\xi\rangle=\left(\begin{array}{c}v_{\xi}+\delta v_{\xi_1}\\
v_{\xi}+\delta v_{\xi_2}\\
v_{\xi}
\end{array}\right),\\
\label{37}&&\langle\phi\rangle=\left(\begin{array}{c} \delta
v_{\phi_1}\\
v_{\phi}+\delta v_{\phi_2}\\
-v_{\phi}
\end{array}\right),~~~~~~~\langle\eta\rangle=\left(\begin{array}{c}\delta v_{\eta_1}\\
v_{\eta}\\
\delta v_{\eta_3}
\end{array}\right)
\end{eqnarray}
where $v_{\xi}$, $v_{\phi}$ and $v_{\eta}$ remain undetermined, and
all the shifts are of order $\varepsilon'^2$ with respect to the LO
VEVs, as is shown in Appendix B. Moreover, all components of
$\langle\chi\rangle$, $\langle\xi\rangle$, $\langle\phi\rangle$ and
$\langle\eta\rangle$ receive different corrections so that the LO
alignment is tilted.
\subsection{Corrections to the charged lepton mass matrix}
In the charged lepton sector, $w_{\ell}$ is corrected by the
following operators
\begin{equation}
\label{38}e^{c}\ell\Phi^3_{\ell}\Phi^2_{\nu}h_d/\Lambda^5,~~~\mu^{c}\ell\Phi^2_{\ell}\Phi^2_{\nu}h_d/\Lambda^4,~~~\tau^{c}\ell\Phi_{\ell}\Phi^2_{\nu}h_d/\Lambda^3
\end{equation}
where all possible contractions among fields are understood. After
lengthy and tedious calculations, we find that each element of
charged lepton mass matrix gets a small correction. Concretely the
corrections to the $e$ row, $\mu$ row and $\tau$ row are of order
$\varepsilon^3\varepsilon'^2v_{d}$,
$\varepsilon^2\varepsilon'^2v_{d}$ and
$\varepsilon\varepsilon'^2v_{d}$ respectively. As a result, the
charged lepton mass matrix with subleading corrections can be
parameterized as
\begin{equation}
\label{39}m_{\ell}=\left(\begin{array}{ccc}
y_e\varepsilon^2 &
y_e\varepsilon^2  &  y_e\varepsilon^2  \\
y_{\mu}\varepsilon &
\omega^2y_{\mu}\varepsilon  &  \omega y_{\mu}\varepsilon  \\
y_{\tau} & \omega y_{\tau}  &  \omega^2y_{\tau}
\end{array}\right)\varepsilon
v_d+\left(\begin{array}{ccc}
a^{\ell}_{11}\varepsilon^2 & a^{\ell}_{12}\varepsilon^2  & a^{\ell}_{13}\varepsilon^2\\
a^{\ell}_{21}\varepsilon & \omega^2a^{\ell}_{22}\varepsilon  & \omega a^{\ell}_{23}\varepsilon\\
a^{\ell}_{31} & \omega a^{\ell}_{32}  & \omega^2a^{\ell}_{33}
\end{array}\right)\varepsilon\varepsilon'^{\,2}v_d
\end{equation}
where the first term denotes the LO contributions, and the second
term represents the corrections induced by the higher dimensional
operators in Eq.(\ref{38}). The coefficients
$a^{\ell}_{ij}(i,j=1,2,3)$ are complex numbers with absolute value
of order one, their specific values are not determined by the flavor
symmetry. Furthermore, we have to consider the corrections from the
shifted vacuum alignment. Since the shifts $\delta v_{\chi_i}$ and
$\delta v_{\xi_i}$ are of order $\varepsilon'^{\,2}v_{\chi}$ and
$\varepsilon'^{\,2}v_{\xi}$ respectively, and the corrections to
each matrix element contain one additional power of $\delta
v_{\chi_i}/v_{\chi}$ or $\delta v_{\xi_i}/v_{\xi}$. Consequently,
including these corrections only amounts to a redefinition of the
$a^{\ell}_{ij}$ parameter in Eq.(\ref{39}). As a result, the unitary
matrix $U_{\ell}$ \footnote{$U_{\ell}$ is the unitary matrix which
diagonalizes the hermitian matrix $m^{\dagger}_{\ell}m_{\ell}$.},
which corresponds to the transformation of the charged leptons used
to diagonalized $m_{\ell}$, is modified into
\begin{equation}
\label{40}U_{\ell}=\frac{1}{\sqrt{3}}\left(\begin{array}{ccc}
1&1&1\\
1&\omega&\omega^2\\
1&\omega^2& \omega
\end{array}\right)U'_{\ell}
\end{equation}
where $U'_{\ell}$ is given by  
\begin{equation}
\label{41}U'_{\ell}=\left(\begin{array}{ccc}1 &
(A_{\ell}\varepsilon'^{2})^{*} &
(B_{\ell}\varepsilon'^{2})^{*}\\
-A_{\ell}\varepsilon'^{2}& 1 & (C_{\ell}\varepsilon'^{2})^{*}\\
-B_{\ell}\varepsilon'^{2} & -C_{\ell}\varepsilon'^{2} &1
\end{array}\right)
\end{equation}
with
\begin{eqnarray}
\nonumber&&A_{\ell}=(a^{\ell}_{21}+\omega^2a^{\ell}_{22}+\omega
a^{\ell}_{23})/(3y_{\mu})\\
\nonumber&&B_{\ell}=(a^{\ell}_{31}+\omega a^{\ell}_{32}+\omega^2
a^{\ell}_{33})/(3y_{\tau})\\
\label{42}&&C_{\ell}=(a^{\ell}_{31}+\omega^2a^{\ell}_{32}+\omega
a^{\ell}_{33})/(3y_{\tau})
\end{eqnarray}
The charged lepton masses are corrected by terms of relative order
$\varepsilon'^{2}$ with respect to LO result, therefore the charged
lepton mass hierarchies predicted at LO are not spoiled by
subleading corrections.
\subsection{Corrections to the neutrino mass matrix}
First of all we focus on the corrections to the right-handed
Majorana neutrino mass. We note that the modified vacuum alignment
doesn't affect the Majorana mass at all, since flavon fields are not
involved in the LO Majorana mass terms. The subleading corrections
from higher dimensional operators are of the form
$\nu^{c}_i\nu^{c}_j\Phi^4_{\nu}/\Lambda^3$, thus every entry of
right-handed Majorana neutrino mass matrix receives corrections of
order $\varepsilon'^{4}\Lambda$ instead of
$\varepsilon'^{2}\Lambda$, which can be safely neglected. Then we
move to consider the corrections to the neutrino Dirac mass. Among
the independent terms of the type
$\nu^{c}_i\ell\Phi^3_{\nu}h_u/\Lambda^3$, only the operators
$(\nu^{c}_i\ell\phi^3h_{u})_{\mathbf{1_1}}/\Lambda^3$ give non-zero
contributions. As a consequence, the first and the third columns of
Dirac mass matrix receive corrections of order
$\varepsilon'^{3}v_u$. In addition to this correction, inserting the
VEV shifts in the LO operators introduces independent corrections of
order $\varepsilon'^{3}v_u$ to the first and second column elements
of the Dirac mass matrix. Including the above two kinds of
corrections mentioned, we conclude that all the elements of Dirac
mass matrix are corrected by terms of ${\cal
O}(\varepsilon'^{3}v_u)$. With these results, we find that each
entry of $m^{SS}_{\nu}$ except the $(11)$ element receives
corrections of order $\varepsilon'^{4}v^2_u/\Lambda$. Now we discuss
the corrections to the Weinberg operators. The superpotential
$w^{eff}_{\nu}$ is corrected by the contraction
\begin{equation}
\label{43}(\ell h_u\ell h_u\Phi^4_{\nu})_{\mathbf{1_1}}/\Lambda^5
\end{equation}
Taking into account the contributions of the modified vacuum
alignment in addition, we find all the elements of $m^{eff}_{\nu}$
receive corrections of order $\varepsilon'^{4}v_u/\Lambda$. As a
result, the overall correction to the light neutrino mass matrix is
a most general symmetric matrix of order
$\varepsilon'^{4}v_u/\Lambda$. The neutrino mass matrix including
subleading corrections can be written as
\begin{equation}
\label{44}m_{\nu}=\varepsilon'^{2}\left(\begin{array}{ccc} x&0& 0\\
0 & y & z\\
0 & z & y
\end{array}\right)\frac{v^2_{u}}{\Lambda}+\varepsilon'^{4}\left(\begin{array}{ccc}
a^{\nu}_{11} &  a^{\nu}_{12}  & a^{\nu}_{13}\\
a^{\nu}_{12}  & a^{\nu}_{22}  &  a^{\nu}_{23}\\
a^{\nu}_{13}  & a^{\nu}_{23}  & a^{\nu}_{33}
\end{array}\right)\frac{v^2_u}{\Lambda}
\end{equation}
where the parameters $x$, $y$ and $z$ can be easily reconstructed
from the LO couplings in Eq.(\ref{13}), and the coefficients
$a^{\nu}_{ij}$ are ${\cal O}(1)$ unspecified constants. The matrix
$m_{\nu}$ is diagonalized by the unitary transformation
\begin{equation}
\label{45}U_{\nu}=\left(\begin{array}{ccc}
0&1&0\\
\frac{1}{\sqrt{2}} & 0 & \frac{i}{\sqrt{2}}\\
\frac{1}{\sqrt{2}} &  0 & \frac{-i}{\sqrt{2}}
\end{array}\right)U'_{\nu}
\end{equation}
where $U'_{\nu}$ is close to an identity matrix with small
corrections on off-diagonal elements, it is given by
\begin{equation}
\label{46}U'_{\nu}=\left(\begin{array}{ccc} 1  &
A_{\nu}\varepsilon'^{2} & B_{\nu}\varepsilon'^{2}\\
-(A_{\nu}\varepsilon'^{2})^{*} & 1 & C_{\nu}\varepsilon'^{2}\\
-(B_{\nu}\varepsilon'^{2})^{*} & -(C_{\nu}\varepsilon'^{2})^{*} &1
\end{array}\right)
\end{equation}
with
\begin{eqnarray}
\nonumber&&A_{\nu}=\frac{(y^{*}+z^{*})(a^{\nu}_{12}+a^{\nu}_{13})+x(a^{\nu*}_{12}+a^{\nu*}_{13})}{\sqrt{2}\,[|x|^2-|y+z|^2]}\\
\nonumber&&B_{\nu}=\frac{i(y^{*}+z^{*})(a^{\nu}_{22}-a^{\nu}_{33})+i(y-z)(a^{\nu*}_{22}-a^{\nu*}_{33})}{-4(yz^{*}+y^{*}z)}\\
\label{47}&&C_{\nu}=\frac{ix^{*}(a^{\nu}_{12}-a^{\nu}_{13})+i(y-z)(a^{\nu*}_{12}-a^{\nu*}_{13})}{\sqrt{2}\,[|y-z|^2-|x|^2]}
\end{eqnarray}
The PMNS matrix is $U_{PMNS}=U^{\dagger}_{\ell}U_{\nu}$, then the
parameters of the lepton mixing matrix are modified as
\begin{eqnarray}
\nonumber&&\sin\theta_{13}=\Big|\frac{1}{\sqrt{3}}(\sqrt{2}B_{\nu}+C_{\nu})\varepsilon'^{2}
-\frac{1}{\sqrt{2}}(A^{*}_{\ell}-B^{*}_{\ell})(\varepsilon'^{2})^{*}\Big|\\
\nonumber&&\sin^2\theta_{12}=\frac{1}{3}+[\frac{1}{3}(\sqrt{2}A_{\nu}-A_{\ell}-B_{\ell})\varepsilon'^{2}+c.c.]\\
\label{48}&&\sin^2\theta_{23}=\frac{1}{2}+[(-\frac{1}{2\sqrt{3}}B_{\nu}+\frac{1}{\sqrt{6}}C_{\nu}+\frac{1}{2}C_{\ell})\varepsilon'^{2}+c.c.]
\end{eqnarray}
We see that all the three mixing angles receive corrections of order
$\varepsilon'^{2}$ from both the neutrino and the charged lepton
sectors. As is pointed out in section 4, the data on solar neutrino
mixing angle $\theta_{12}$ constrain $\varepsilon'\leq\lambda_c$.
Then the reactor angle $\theta_{13}$ is of order $\lambda^2_c$, it
is within the sensitivity of the experiments which are now in
preparation and will take data in the near future \cite{Ardellier:2006mn,Wang:2006ca}. Since three complex parameters which
are related with three light neutrino masses are involved at LO, the light neutrino mass spectrum can be normal hierarchy or inverted hierarchy,
and the phenomenological predictions of the model are just the generic results of neutrino
mass matrix with TB mixing, e.g., degenerate neutrino mass spectrum is disfavored since strong fine-tuning is required to produce the observed mass squared differences
$\Delta m^2_{sol}$ and $\Delta m^2_{atm}$, and the $0\nu2\beta$-decay mass
$|m_{ee}|$ in inverted hierarchy is generally larger than that in normal hierarchy.

\section{Phenomenological implications}
\begin{figure}[hptb]
\begin{center}
\includegraphics[scale=.56]{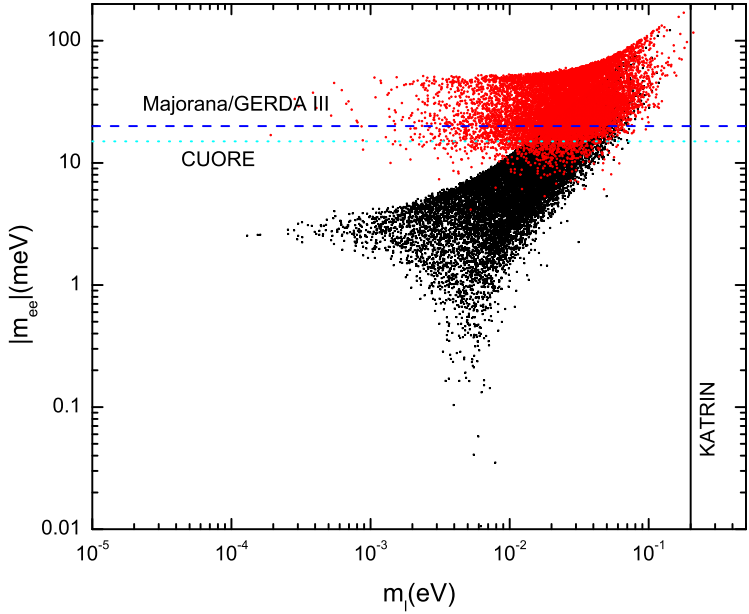}
\end{center}
\caption{\label{fig:mee}The effective mass $|m_{ee}|$ as a function
of the lightest neutrino mass. The red corresponds to the inverted
hierarchy neutrino mass spectrum, and the black corresponds to the
normal hierarchy case. The future sensitivity of 0.2 eV of KATRIN
experiment is shown by the vertical solid line, the future expected
bounds on $|m_{ee}|$ from CUORE and Majorana/GERDA III experiments
are represented by horizontal lines.}
\end{figure}

In the following, we shall investigate the physical consequences of
our model, and the corresponding predictions are presented. We
perform a numerical analysis by treating all the LO and NLO
coefficients as random complex numbers with absolute value between
1/3 and 3, the expansion parameter $\varepsilon$ and $\varepsilon'$
are set to the indicative values 0.05 and 0.22 respectively. In
Fig.\ref{fig:mee}, we plot the effective $0\nu2\beta$-decay mass
$|m_{ee}|$ as a function of the lightest neutrino mass. The
constraints which have been imposed to draw the points are the
experimental values at $3\sigma$ for the neutrino oscillation
parameters $\Delta m^2_{atm}$, $\Delta m^2_{sol}$,
$\sin^2\theta_{12}$, $\sin^2\theta_{23}$ and $\sin^2\theta_{13}$
\cite{Schwetz:2008er,Fogli:Indication,GonzalezGarcia:2010er}. We
also show the future sensitivity on the lightest neutrino mass of
0.2 eV from the KATRIN experiment \cite{katrin}, and the horizontal
lines represent the sensitivities of the future $0\nu2\beta$-decay
experiments CUORE \cite{cuore} and Majorana \cite{majorana}/GERDA
III \cite{gerda}, which are 15 meV and 20 meV respectively. It is
obvious that the effective mass $|m_{ee}|$ of inverted hierarchy
(IH) is generally larger than that of normal hierarchy (NH). Since
the bulk of data are predicted to be above the sensitivity of CUORE
experiment for IH, the rare process $0\nu2\beta$-decay should be
observable in future, if the neutrino spectrum is IH. We note that
most of the points fall into the region where the lightest neutrino
mass is smaller than 0.02 eV for NH spectrum, and a large set of
points lie in the region of the lightest neutrino mass between 0.01
eV and 0.04 eV for IH case. The values beyond these regions, in
particular the region of degenerate spectrum, are strongly
disfavored.

\begin{figure}[hptb]
\begin{center}
\includegraphics[scale=.56]{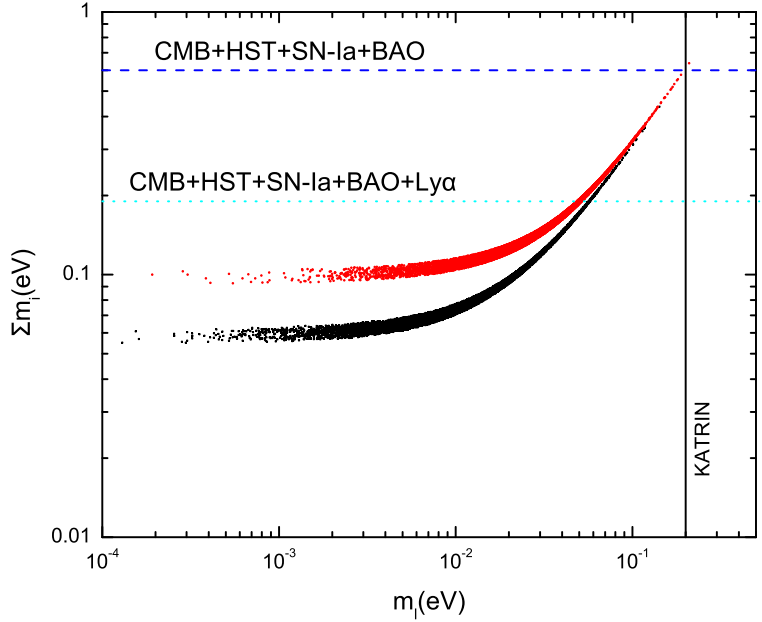}
\end{center}
\caption{\label{fig:Mnu_sum}The effective mass $|m_{ee}|$ as a
function of the lightest neutrino mass. The red corresponds to the
inverted hierarchy neutrino mass spectrum, and the black corresponds
to the normal hierarchy case. The vertical solid line represents the
future sensitivity of 0.2 eV from the KATRIN experiment, and the
horizontal lines refer to the cosmological bounds.}
\end{figure}

Finally, we show the sum of light neutrino mass as a function of the
lightest neutrino mass in Fig.\ref{fig:Mnu_sum}. The vertical line
denotes the future sensitivity of KATRIN experiment, and the
horizontal lines are the cosmological bounds \cite{Fogli:2006yq}.
The first one is at 0.60 eV, which is obtained by combining the data
in Ref.\cite{cosmo}, and the second one at 0.19 eV corresponds to
all the previous data combined to the small scale primordial
spectrum from Lyman-alpha (Ly$\alpha$) forest clouds \cite{Ly}. We
see that the current cosmological information on the sum of the
neutrino masses can hardly distinguish the NH spectrum from the IH
spectrum. However, such a discrimination could be possible if these
bounds are improved in the near future.

\section{Conclusions and discussions}
In this work, we have presented a $T_{13}$ model for TB mixing based
on the flavor symmetry $T_{13}\times Z_4\times Z_2$. Both the
charged lepton singlets $e^{c}$, $\mu^{c}$, $\tau^{c}$ and the
right-handed neutrinos $\nu^{c}_i$ are assigned as $T_{13}$ singlets
in this work. The light neutrino masses are generated as a
combination of type I see-saw mechanism and Weinberg operators, and
neutrino mass spectrum can be normal hierarchy or inverted
hierarchy. In the charged lepton sector, the flavon fields
$\Phi_{\ell}=\{\chi,\xi\}$ break the $T_{13}$ group into the $Z_3$
subgroup at LO, and the symmetry breaking parameter
$\varepsilon\equiv\langle\chi\rangle/\Lambda\sim\langle\xi\rangle/\Lambda$
controls the charged lepton mass hierarchies without invoking a
Froggatt-Nielsen $U(1)$ symmetry. In the neutrino sector, the
$T_{13}$ group is entirely broken by the flavon fields
$\Phi_{\nu}=\{\phi,\eta\}$ at LO, the symmetry breaking parameter
$\varepsilon'\equiv\langle\phi\rangle/\Lambda\sim\langle\eta\rangle/\Lambda$
can be chosen to be of the order of Cabibbo angle $\lambda_c$
without spoiling the LO predictions and vacuum alignment. It is a
noticeable feature that we can still reproduce the TB mixing even if
the flavor symmetry is broken to nothing in the neutrino sector at
LO.

The subleading corrections are discussed in detail. The subleading
operators contributing to lepton mass and vacuum alignment are
obtained by inserting $\Phi^2_{\nu}$ into the LO operators in all
possible ways and by extracting the $T_{13}\times Z_4\times Z_2$
invariants. We have showed that all the mixing angles receive
corrections at the level of ${\cal O}(\lambda^2_c)$, in particular,
the reactor angle $\theta_{13}$ is predicted to be within the reach
of next generation neutrino oscillation experiments, although it is
small. Furthermore, since the neutrino mass matrix depends on three unrelated complex parameters at LO,
the phenomenological consequences of the model are the general results of neutrino mass matrix with TB mixing,
there are no model-dependent peculiar predictions.

In the end, we discuss whether we can extend the $T_{13}$ flavor
symmetry from the neutrino sector to the quark sector. The most
naive way is to adopt for quarks the same classification scheme
under $T_{13}\times Z_4\times Z_2$ that we have used for leptons.
With such an assignment, both up and down type quark mass matrices
are diagonalized by the same unitary matrix $U_{\ell}$ shown in
Eq.(\ref{9}), as a consequence the CKM matrix is a unit matrix at
LO, this is a good first order approximation. The off-diagonal
elements of the CKM matrix arise when the subleading contributions
are taken into account. As has been showed in section 5, the
subleading corrections to the three quark mixing angles are of order
$\lambda^2_c$, the resulting CKM matrix should have the same form of
the unitary matrix $U'_{\ell}$ given in Eq.(\ref{41}). Therefore it
seems difficult to reproduce the quark mixing without introducing
new ingredients in the symmetry breaking sector. Furthermore, there
is another lack if we adopt for quark the same structure as that in
the lepton sector, the resulting mass hierarchies among the up type
quarks are not realistic, although it is a satisfactory result that
the mass spectrums of down type quarks and charged leptons are
predicted to have the same pattern. Since the top quark mass is of
order of the electroweak symmetry breaking scale, it is much heavier
than the remaining quarks, it is natural to assign quarks as 2+1
representation instead of a triplet. In the context of U(2) flavor
group, this assignment has been known to give realistic quark mixing
matrix and mass hierarchies \cite{u2}. Inspired by this assignment,
it is usually suggested to extend the flavor symmetry group, which
can produce the neutrino TB mixing, to its double covering in order
to give a coherent description of all fermion masses and flavor
mixings. The flavor models based on $T^{\,\prime}$ and ${\cal
I}^{\,\prime}$ groups \cite{Tprime,A5prime}, which are the double
covering groups of $A_4$ and $A_5$ respectively, have been studied
extensively. These models can really lead to a good description of
the observed pattern of quark masses and mixing besides reproducing
TB mixing (or the gold ratio mixing pattern) for neutrinos.
Following the same logic, we expect the double covering group of
$T_{13}$ can simultaneously describe the lepton and quark sector
very well.

\section*{Acknowledgements}

This work is supported by the National Natural Science Foundation of
China under Grant No.10905053, Chinese Academy KJCX2-YW-N29 and the
973 project with Grant No. 2009CB825200.

\vfill
\newpage

\section*{\label{app:A}Appendix A: Representation matrices and Clebsch-Gordan coefficients of $T_{13}$}
The $T_{13}$ group has 7 inequivalent irreducible representations
$\mathbf{1_1}$, $\mathbf{1_2}$, $\mathbf{1_3}$, $\mathbf{3_1}$,
$\mathbf{\bar{3}_1}$, $\mathbf{3_2}$ and $\mathbf{\bar{3}_2}$. The
representation matrices of the generators $S$ and $T$ in these
representations are given in Eq.(\ref{1}) and Eq.(\ref{2}). In the
following, we present the representation matrices of all the group
elements for the three dimensional representations. The explicit
expressions of the elements in the $3_1$ representation are
\begin{eqnarray}
\nonumber\hspace{-11.3cm}{\cal C}_1:&&~e=\left(\begin{array}{ccc} 1&0&0\\
0&1&0\\
0&0&1
\end{array}\right)
\end{eqnarray}
\begin{eqnarray}
\nonumber {\cal C}_2:&&~T=\left(\begin{array}{ccc} 0&0&1\\
1&0&0\\
0&1&0
\end{array}\right),~~TS=\left(\begin{array}{ccc}
0&0&\rho^9\\
\rho&0&0\\
0&\rho^3&0
\end{array}\right),~~TS^2=\left(\begin{array}{ccc}
0&0&\rho^5\\
\rho^2&0&0\\
0&\rho^6&0
\end{array}\right),\\
\nonumber&&TS^3=\left(\begin{array}{ccc}
0&0&\rho\\
\rho^3&0&0\\
0&\rho^9&0
\end{array}\right),~~~TS^4=\left(\begin{array}{ccc}
0&0&\rho^{10}\\
\rho^4&0&0\\
0&\rho^{12}&0
\end{array}\right),~~TS^5=\left(\begin{array}{ccc}
0&0&\rho^6\\
\rho^5&0&0\\
0&\rho^2&0
\end{array}\right)\\
\nonumber&&TS^6=\left(\begin{array}{ccc}
0&0&\rho^2\\
\rho^6&0&0\\
0&\rho^5&0
\end{array}\right),~~TS^7=\left(\begin{array}{ccc}
0&0&\rho^{11}\\
\rho^7&0&0\\
0&\rho^8&0
\end{array}\right),~~TS^8=\left(\begin{array}{ccc}
0&0&\rho^7\\
\rho^8&0&0\\
0&\rho^{11}&0
\end{array}\right)\\
\nonumber&&TS^9=\left(\begin{array}{ccc}
0&0&\rho^3\\
\rho^9&0&0\\
0&\rho&0
\end{array}\right),~~TS^{10}=\left(\begin{array}{ccc}
0&0&\rho^{12}\\
\rho^{10}&0&0\\
0&\rho^4&0
\end{array}\right),~~TS^{11}=\left(\begin{array}{ccc}
0&0&\rho^8\\
\rho^{11}&0&0\\
0&\rho^7&0
\end{array}\right),\\
\nonumber&&TS^{12}=\left(\begin{array}{ccc}
0&0&\rho^4\\
\rho^{12}&0&0\\
0&\rho^{10}&0
\end{array}\right)
\end{eqnarray}
\begin{eqnarray}
\nonumber {\cal C}_3:&&~T^2=\left(\begin{array}{ccc} 0&1&0\\
0&0&1\\
1&0&0
\end{array}\right),~~T^2S=\left(\begin{array}{ccc} 0&\rho^3&0\\
0&0&\rho^9\\
\rho&0&0
\end{array}\right),~~T^2S^2=\left(\begin{array}{ccc} 0&\rho^6&0\\
0&0&\rho^5\\
\rho^2&0&0
\end{array}\right),\\
\nonumber&&~~T^2S^3=\left(\begin{array}{ccc} 0&\rho^9&0\\
0&0&\rho\\
\rho^3&0&0
\end{array}\right),~~T^2S^4=\left(\begin{array}{ccc} 0&\rho^{12}&0\\
0&0&\rho^{10}\\
\rho^4&0&0
\end{array}\right),~~T^2S^5=\left(\begin{array}{ccc} 0&\rho^2&0\\
0&0&\rho^6\\
\rho^5&0&0
\end{array}\right),\\
\nonumber&&~T^2S^6=\left(\begin{array}{ccc} 0&\rho^5&0\\
0&0&\rho^2\\
\rho^6&0&0
\end{array}\right),~~T^2S^7=\left(\begin{array}{ccc} 0&\rho^8&0\\
0&0&\rho^{11}\\
\rho^{7}&0&0
\end{array}\right),~~T^2S^8=\left(\begin{array}{ccc} 0&\rho^{11}&0\\
0&0&\rho^7\\
\rho^8&0&0
\end{array}\right),\\
\nonumber&&~T^2S^9=\left(\begin{array}{ccc} 0&\rho&0\\
0&0&\rho^3\\
\rho^9&0&0
\end{array}\right),~~T^2S^{10}=\left(\begin{array}{ccc} 0&\rho^4&0\\
0&0&\rho^{12}\\
\rho^{10}&0&0
\end{array}\right),~~T^2S^{11}=\left(\begin{array}{ccc} 0&\rho^7&0\\
0&0&\rho^8\\
\rho^{11}&0&0
\end{array}\right),\\
\nonumber&&~T^2S^{12}=\left(\begin{array}{ccc} 0&\rho^{10}&0\\
0&0&\rho^4\\
\rho^{12}&0&0
\end{array}\right)
\end{eqnarray}
\begin{eqnarray}
\nonumber{\cal C}_4:&&~S=\left(\begin{array}{ccc} \rho& 0&0\\
0&\rho^3&0\\
0&0&\rho^9
\end{array}\right),~~S^3=\left(\begin{array}{ccc} \rho^3& 0&0\\
0&\rho^9&0\\
0&0&\rho
\end{array}\right),~~S^9=\left(\begin{array}{ccc} \rho^9& 0&0\\
0&\rho &0\\
0&0&\rho^3
\end{array}\right)\\
\nonumber {\cal C}_5:&&~S^4=\left(\begin{array}{ccc} \rho^4& 0&0\\
0&\rho^{12}&0\\
0&0&\rho^{10}
\end{array}\right),~~S^{10}=\left(\begin{array}{ccc} \rho^{10}& 0&0\\
0&\rho^4&0\\
0&0&\rho^{12}
\end{array}\right),~~S^{12}=\left(\begin{array}{ccc} \rho^{12}& 0&0\\
0&\rho^{10}&0\\
0&0&\rho^4
\end{array}\right)\\
\nonumber {\cal C}_6:&&~S^2=\left(\begin{array}{ccc} \rho^2& 0&0\\
0&\rho^6&0\\
0&0&\rho^5
\end{array}\right),~~S^5=\left(\begin{array}{ccc} \rho^5& 0&0\\
0&\rho^2&0\\
0&0&\rho^6
\end{array}\right),~~S^6=\left(\begin{array}{ccc} \rho^6& 0&0\\
0&\rho^5&0\\
0&0&\rho^2
\end{array}\right)\\
\nonumber {\cal C}_7:&&~S^7=\left(\begin{array}{ccc} \rho^7& 0&0\\
0&\rho^8&0\\
0&0&\rho^{11}
\end{array}\right),~~S^8=\left(\begin{array}{ccc} \rho^8& 0&0\\
0&\rho^{11}&0\\
0&0&\rho^7
\end{array}\right),~~S^{11}=\left(\begin{array}{ccc} \rho^{11}& 0&0\\
0&\rho^7&0\\
0&0&\rho^8
\end{array}\right)
\end{eqnarray}
while for the 3-dimensional representation $\mathbf{3_2}$ the
elements are
\begin{eqnarray}
\nonumber\hspace{-11.3cm} {\cal
C}_1:&&~e=\left(\begin{array}{ccc}1&0&0\\
0&1&0\\
0&0&1
\end{array}\right)
\end{eqnarray}
\begin{eqnarray}
\nonumber {\cal C}_2:&&~T=\left(\begin{array}{ccc} 0&0&1\\
1&0&0\\
0&1&0
\end{array}\right),~~TS=\left(\begin{array}{ccc} 0&0&\rho^5\\
\rho^2&0&0\\
0&\rho^6&0
\end{array}\right),~~TS^2=\left(\begin{array}{ccc} 0&0&\rho^{10}\\
\rho^4&0&0\\
0&\rho^{12}&0
\end{array}\right),\\
\nonumber&&~TS^3=\left(\begin{array}{ccc} 0&0&\rho^2\\
\rho^6&0&0\\
0&\rho^5&0
\end{array}\right),~~TS^4=\left(\begin{array}{ccc} 0&0&\rho^7\\
\rho^8&0&0\\
0&\rho^{11}&0
\end{array}\right),~~TS^5=\left(\begin{array}{ccc} 0&0&\rho^{12}\\
\rho^{10}&0&0\\
0&\rho^4&0
\end{array}\right),\\
\nonumber&&~TS^6=\left(\begin{array}{ccc} 0&0&\rho^4\\
\rho^{12}&0&0\\
0&\rho^{10}&0
\end{array}\right),~~TS^7=\left(\begin{array}{ccc} 0&0&\rho^9\\
\rho&0&0\\
0&\rho^3&0
\end{array}\right),~~TS^8=\left(\begin{array}{ccc} 0&0&\rho\\
\rho^3&0&0\\
0&\rho^9&0
\end{array}\right),\\
\nonumber&&~~TS^9=\left(\begin{array}{ccc} 0&0&\rho^6\\
\rho^5&0&0\\
0&\rho^2&0
\end{array}\right),~~TS^{10}=\left(\begin{array}{ccc} 0&0&\rho^{11}\\
\rho^7&0&0\\
0&\rho^8&0
\end{array}\right),~~TS^{11}=\left(\begin{array}{ccc} 0&0&\rho^3\\
\rho^9&0&0\\
0&\rho&0
\end{array}\right),\\
\nonumber&&TS^{12}=\left(\begin{array}{ccc} 0&0&\rho^8\\
\rho^{11}&0&0\\
0&\rho^7&0
\end{array}\right)
\end{eqnarray}
\begin{eqnarray}
\nonumber {\cal C}_3:&&~~~T^2=\left(\begin{array}{ccc} 0&1&0\\
0&0&1\\
1&0&0
\end{array}\right),~~T^2S=\left(\begin{array}{ccc} 0&\rho^6 &0\\
0&0&\rho^5\\
\rho^2&0&0
\end{array}\right),~~T^2S^2=\left(\begin{array}{ccc} 0&\rho^{12} &0\\
0&0&\rho^{10}\\
\rho^4&0&0
\end{array}\right),\\
\nonumber&&T^2S^3=\left(\begin{array}{ccc} 0&\rho^5 &0\\
0&0&\rho^2\\
\rho^6&0&0
\end{array}\right),~~T^2S^4=\left(\begin{array}{ccc} 0&\rho^{11} &0\\
0&0&\rho^7\\
\rho^8&0&0
\end{array}\right),~~T^2S^5=\left(\begin{array}{ccc} 0&\rho^4 &0\\
0&0&\rho^{12}\\
\rho^{10}&0&0
\end{array}\right),\\
\nonumber&&T^2S^6=\left(\begin{array}{ccc} 0&\rho^{10} &0\\
0&0&\rho^4\\
\rho^{12}&0&0
\end{array}\right),~~T^2S^7=\left(\begin{array}{ccc} 0&\rho^3 &0\\
0&0&\rho^9\\
\rho&0&0
\end{array}\right),~~T^2S^8=\left(\begin{array}{ccc} 0&\rho^9 &0\\
0&0&\rho\\
\rho^3&0&0
\end{array}\right),\\
\nonumber&&~T^2S^9=\left(\begin{array}{ccc} 0&\rho^2 &0\\
0&0&\rho^6\\
\rho^5&0&0
\end{array}\right),~~T^2S^{10}=\left(\begin{array}{ccc} 0&\rho^8 &0\\
0&0&\rho^{11}\\
\rho^7&0&0
\end{array}\right),~~T^2S^{11}=\left(\begin{array}{ccc} 0&\rho &0\\
0&0&\rho^3\\
\rho^9&0&0
\end{array}\right),\\
\nonumber&&T^2S^{12}=\left(\begin{array}{ccc} 0&\rho^7 &0\\
0&0&\rho^8\\
\rho^{11}&0&0
\end{array}\right)
\end{eqnarray}
\begin{eqnarray}
\nonumber {\cal
C}_4:&&~S=\left(\begin{array}{ccc}\rho^2&0&0\\
0&\rho^6&0\\
0&0&\rho^5
\end{array}\right),~~S^3=\left(\begin{array}{ccc}\rho^6&0&0\\
0&\rho^5&0\\
0&0&\rho^2
\end{array}\right),~~S^9=\left(\begin{array}{ccc}\rho^5&0&0\\
0&\rho^2&0\\
0&0&\rho^6
\end{array}\right)\\
\nonumber {\cal
C}_5:&&~S^4=\left(\begin{array}{ccc}\rho^8&0&0\\
0&\rho^{11}&0\\
0&0&\rho^7
\end{array}\right),~~S^{10}=\left(\begin{array}{ccc}\rho^7&0&0\\
0&\rho^8&0\\
0&0&\rho^{11}
\end{array}\right),~~S^{12}=\left(\begin{array}{ccc}\rho^{11}&0&0\\
0&\rho^7&0\\
0&0&\rho^8
\end{array}\right)\\
\nonumber {\cal
C}_6:&&~S^2=\left(\begin{array}{ccc}\rho^4&0&0\\
0&\rho^{12}&0\\
0&0&\rho^{10}
\end{array}\right),~~S^5=\left(\begin{array}{ccc}\rho^{10}&0&0\\
0&\rho^4&0\\
0&0&\rho^{12}
\end{array}\right),~~S^6=\left(\begin{array}{ccc}\rho^{12}&0&0\\
0&\rho^{10}&0\\
0&0&\rho^4
\end{array}\right)\\
\nonumber {\cal
C}_7:&&~S^7=\left(\begin{array}{ccc}\rho&0&0\\
0&\rho^3&0\\
0&0&\rho^9
\end{array}\right),~~S^8=\left(\begin{array}{ccc}\rho^3&0&0\\
0&\rho^9&0\\
0&0&\rho
\end{array}\right),~~S^{11}=\left(\begin{array}{ccc}\rho^9&0&0\\
0&\rho&0\\
0&0&\rho^3
\end{array}\right)
\end{eqnarray}
For the remaining 3-dimensional representations $\mathbf{\bar{3}_1}$
and $\mathbf{\bar{3}_2}$, the matrices representing the elements of
the group can be found from those just listed for the
representations $\mathbf{3_1}$ and $\mathbf{3_2}$ by performing
complex conjugation. The above representation matrices can help us
to see clearly how the $T_{13}$ flavor symmetry is broken in model
building. Starting from the above explicit representation matrices,
we can straightforwardly get the product decomposition rules of the
$T_{13}$ group. In the following we use $\alpha_i$ to denote the
elements of the first representation of the product and $\beta_i$ to
indicate those of the second representation.

\begin{itemize}

\item{$\mathbf{1_2}\otimes\mathbf{3_1}=\mathbf{3_1}$}
\begin{equation}
\mathbf{3_1}\sim\left(\begin{array}{c} \alpha\beta_1\\
\omega^2\alpha\beta_2\\
\omega\alpha\beta_3
\end{array}\right)
\end{equation}

\item{$\mathbf{1_2}\otimes\mathbf{\overline{3}_1}=\mathbf{\overline{3}_1}$}
\begin{equation}
\overline{3}_1\sim\left(\begin{array}{c} \alpha\beta_1\\
\omega^2\alpha\beta_2\\
\omega\alpha\beta_3
\end{array}\right)
\end{equation}

\item{$\mathbf{1}_2\otimes\mathbf{3}_2=\mathbf{3}_2$}
\begin{equation}
3_2\sim\left(\begin{array}{c} \alpha\beta_1\\
\omega^2\alpha\beta_2\\
\omega\alpha\beta_3
\end{array}\right)
\end{equation}

\item{$\mathbf{1_2}\otimes\mathbf{\overline{3}_2}=\mathbf{\overline{3}_2}$}
\begin{equation}
\overline{3}_2\sim\left(\begin{array}{c} \alpha\beta_1\\
\omega^2\alpha\beta_2\\
\omega\alpha\beta_3
\end{array}\right)
\end{equation}

\item{$\mathbf{1_3}\otimes\mathbf{3_1}=\mathbf{3_1}$}
\begin{equation}
\mathbf{3_1}\sim\left(\begin{array}{c} \alpha\beta_1\\
\omega\alpha\beta_2\\
\omega^2\alpha\beta_3
\end{array}\right)
\end{equation}

\item{$\mathbf{1_3}\otimes\mathbf{\overline{3}_1}=\mathbf{\overline{3}_1}$}
\begin{equation}
\mathbf{\overline{3}_1}\sim\left(\begin{array}{c} \alpha\beta_1\\
\omega\alpha\beta_2\\
\omega^2\alpha\beta_3
\end{array}\right)
\end{equation}

\item{$\mathbf{1_3}\otimes\mathbf{3_2}=\mathbf{3_2}$}
\begin{equation}
\mathbf{3_2}\sim\left(\begin{array}{c} \alpha\beta_1\\
\omega\alpha\beta_2\\
\omega^2\alpha\beta_3
\end{array}\right)
\end{equation}

\item{$\mathbf{1_3}\otimes\mathbf{\overline{3}_2}=\mathbf{\overline{3}_2}$}
\begin{equation}
\mathbf{\overline{3}_2}\sim\left(\begin{array}{c} \alpha\beta_1\\
\omega\alpha\beta_2\\
\omega^2\alpha\beta_3
\end{array}\right)
\end{equation}

\item{$\mathbf{3_1}\otimes\mathbf{3_1}=\mathbf{\overline{3}}_{1S}\oplus\mathbf{\overline{3}}_{1A}\oplus\mathbf{3_2}$}
\begin{equation}
\overline{3}_{1S}\sim\left(\begin{array}{c}
\alpha_2\beta_3+\alpha_3\beta_2\\
\alpha_3\beta_1+\alpha_1\beta_3\\
\alpha_1\beta_2+\alpha_2\beta_1
\end{array} \right)
\end{equation}

\begin{equation}
\overline{3}_{1A}\sim\left(\begin{array}{c}
\alpha_2\beta_3-\alpha_3\beta_2\\
\alpha_3\beta_1-\alpha_1\beta_3\\
\alpha_1\beta_2-\alpha_2\beta_1
\end{array} \right)
\end{equation}

\begin{equation}
3_2\sim\left(\begin{array}{c}\alpha_1\beta_1\\
\alpha_2\beta_2\\
\alpha_3\beta_3
\end{array} \right)
\end{equation}

\item{$\mathbf{3_1}\otimes\mathbf{\overline{3}_1}=\mathbf{1}_1\oplus\mathbf{1}_2\oplus\mathbf{1}_3\oplus\mathbf{3}_2\oplus\mathbf{\overline{3}_2}$}
\begin{equation}
\mathbf{1}_1\sim\alpha_1\beta_1+\alpha_2\beta_2+\alpha_3\beta_3
\end{equation}
\begin{equation}
\mathbf{1}_2\sim\alpha_1\beta_1+\omega\alpha_2\beta_2+\omega^2\alpha_3\beta_3
\end{equation}
\begin{equation}
\mathbf{1}_3\sim\alpha_1\beta_1+\omega^2\alpha_2\beta_2+\omega\alpha_3\beta_3
\end{equation}
\begin{equation}
\mathbf{3}_2\sim\left(\begin{array}{c}
\alpha_2\beta_1\\
\alpha_3\beta_2\\
\alpha_1\beta_3\\
\end{array}\right)
\end{equation}
\begin{equation}
\mathbf{\overline{3}}_2\sim\left(\begin{array}{c}
\alpha_1\beta_2\\
\alpha_2\beta_3\\
\alpha_3\beta_1\\
\end{array}\right)
\end{equation}

\item{$\mathbf{3_1}\otimes\mathbf{3_2}=\mathbf{3}_1\oplus\mathbf{3}_2\oplus\mathbf{\overline{3}_2}$}
\begin{equation}
\mathbf{3}_1\sim\left(\begin{array}{c} \alpha_3\beta_3\\
\alpha_1\beta_1\\
\alpha_2\beta_2
\end{array}\right)
\end{equation}
\begin{equation}
\mathbf{3}_2\sim\left(\begin{array}{c}\alpha_3\beta_2\\
\alpha_1\beta_3\\
\alpha_2\beta_1
\end{array}\right)
\end{equation}
\begin{equation}
\mathbf{\overline{3}}_2\sim\left(\begin{array}{c}\alpha_3\beta_1\\
\alpha_1\beta_2\\
\alpha_2\beta_3
\end{array}\right)
\end{equation}

\item{$\mathbf{3_1}\otimes\mathbf{\overline{3}_2}=\mathbf{3}_1\oplus\mathbf{\overline{3}_1}\oplus\mathbf{\overline{3}_2}$}
\begin{equation}
\mathbf{3_1}\sim\left(\begin{array}{c} \alpha_2\beta_1\\
\alpha_3\beta_2\\
\alpha_1\beta_3\\
\end{array}\right)
\end{equation}
\begin{equation}
\mathbf{\overline{3}_1}\sim\left(\begin{array}{c} \alpha_1\beta_1\\
\alpha_2\beta_2\\
\alpha_3\beta_3
\end{array}\right)
\end{equation}
\begin{equation}
\overline{3}_2\sim\left(\begin{array}{c} \alpha_2\beta_3\\
\alpha_3\beta_1\\
\alpha_1\beta_2
\end{array}\right)
\end{equation}

\item{$\mathbf{\overline{3}_1}\otimes\mathbf{\overline{3}_1}=\mathbf{3}_{1S}\oplus\mathbf{3}_{1A}\oplus\mathbf{\overline{3}_2}$}
\begin{equation}
\mathbf{3}_{1S}\sim\left(\begin{array}{c}\alpha_2\beta_3+\alpha_3\beta_2\\
\alpha_3\beta_1+\alpha_1\beta_3\\
\alpha_1\beta_2+\alpha_2\beta_1
\end{array}\right)
\end{equation}
\begin{equation}
\mathbf{3}_{1A}\sim\left(\begin{array}{c}\alpha_2\beta_3-\alpha_3\beta_2\\
\alpha_3\beta_1-\alpha_1\beta_3\\
\alpha_1\beta_2-\alpha_2\beta_1
\end{array}\right)
\end{equation}
\begin{equation}
\mathbf{\overline{3}_2}\sim\left(\begin{array}{c} \alpha_1\beta_1\\
\alpha_2\beta_2\\
\alpha_3\beta_3\\
\end{array}\right)
\end{equation}

\item{$\mathbf{\overline{3}_1}\otimes\mathbf{3_2}=\mathbf{3_1}\oplus\mathbf{\overline{3}_1}\oplus\mathbf{3_2}$}
\begin{equation}
\mathbf{3_1}\sim\left(\begin{array}{c} \alpha_1\beta_1\\
\alpha_2\beta_2\\
\alpha_3\beta_3
\end{array}\right)
\end{equation}
\begin{equation}
\mathbf{\overline{3}_1}\sim\left(\begin{array}{c} \alpha_2\beta_1\\
\alpha_3\beta_2\\
\alpha_1\beta_3\\
\end{array}\right)
\end{equation}
\begin{equation}
\mathbf{3_2}\sim\left(\begin{array}{c}
\alpha_2\beta_3\\
\alpha_3\beta_1\\
\alpha_1\beta_2\\
\end{array}\right)
\end{equation}

\item{$\mathbf{\overline{3}_1}\otimes\mathbf{\overline{3}_2}=\mathbf{\overline{3}_1}\oplus\mathbf{3_2}\oplus\mathbf{\overline{3}_2}$}
\begin{equation}
\mathbf{\overline{3}_1}\sim\left(\begin{array}{c}
\alpha_3\beta_3\\
\alpha_1\beta_1\\
\alpha_2\beta_2\\
\end{array}\right)
\end{equation}
\begin{equation}
\mathbf{3_2}\sim\left(\begin{array}{c}
\alpha_3\beta_1\\
\alpha_1\beta_2\\
\alpha_2\beta_3\\
\end{array}\right)
\end{equation}
\begin{equation}
\mathbf{\overline{3}_2}\sim\left(\begin{array}{c}
\alpha_3\beta_2\\
\alpha_1\beta_3\\
\alpha_2\beta_1\\
\end{array}\right)
\end{equation}

\item{$\mathbf{3_2}\otimes\mathbf{3_2}=\mathbf{\overline{3}_1}\oplus\mathbf{\overline{3}}_{2S}\oplus\mathbf{\overline{3}}_{2A}$}
\begin{equation}
\mathbf{\overline{3}_1}\sim\left(\begin{array}{c}
\alpha_2\beta_2\\
\alpha_3\beta_3\\
\alpha_1\beta_1\\
\end{array}\right)
\end{equation}
\begin{equation}
\mathbf{\overline{3}}_{2S}\sim\left(\begin{array}{c}
\alpha_2\beta_3+\alpha_3\beta_2\\
\alpha_3\beta_1+\alpha_1\beta_3\\
\alpha_1\beta_2+\alpha_2\beta_1\\
\end{array}\right)
\end{equation}
\begin{equation}
\mathbf{\overline{3}}_{2A}\sim\left(\begin{array}{c}
\alpha_2\beta_3-\alpha_3\beta_2\\
\alpha_3\beta_1-\alpha_1\beta_3\\
\alpha_1\beta_2-\alpha_2\beta_1\\
\end{array}\right)
\end{equation}

\item{$\mathbf{3_2}\otimes\mathbf{\overline{3}_2}=\mathbf{1_1}\oplus\mathbf{1_2}\oplus\mathbf{1_3}\oplus\mathbf{3_1}\oplus\mathbf{\overline{3}_1}$}
\begin{equation}
\mathbf{1_1}\sim\alpha_1\beta_1+\alpha_2\beta_2+\alpha_3\beta_3
\end{equation}
\begin{equation}
\mathbf{1_2}\sim\alpha_1\beta_1+\omega\alpha_2\beta_2+\omega^2\alpha_3\beta_3
\end{equation}
\begin{equation}
\mathbf{1_3}\sim\alpha_1\beta_1+\omega^2\alpha_2\beta_2+\omega\alpha_3\beta_3
\end{equation}
\begin{equation}
\mathbf{3_1}\sim\left(\begin{array}{c}\alpha_2\beta_3\\
\alpha_3\beta_1\\
\alpha_1\beta_2\\
\end{array}
\right)
\end{equation}
\begin{equation}
\mathbf{\overline{3}_1}\sim\left(\begin{array}{c}\alpha_3\beta_2\\
\alpha_1\beta_3\\
\alpha_2\beta_1\\
\end{array}
\right)
\end{equation}

\item{$\mathbf{\overline{3}_2}\otimes\mathbf{\overline{3}_2}=\mathbf{3_1}\oplus\mathbf{3}_{2S}\oplus\mathbf{3}_{2A}$}
\begin{equation}
\mathbf{3_1}\sim\left(\begin{array}{c}
\alpha_2\beta_2\\
\alpha_3\beta_3\\
\alpha_1\beta_1\\
\end{array}\right)
\end{equation}
\begin{equation}
\mathbf{3}_{2S}\sim\left(\begin{array}{c}
\alpha_2\beta_3+\alpha_3\beta_2\\
\alpha_3\beta_1+\alpha_1\beta_3\\
\alpha_1\beta_2+\alpha_2\beta_1
\end{array}\right)
\end{equation}
\begin{equation}
\mathbf{3}_{2A}\sim\left(\begin{array}{c}
\alpha_2\beta_3-\alpha_3\beta_2\\
\alpha_3\beta_1-\alpha_1\beta_3\\
\alpha_1\beta_2-\alpha_2\beta_1
\end{array}\right)
\end{equation}

\end{itemize}

\section*{Appendix B: Vacuum alignment beyond leading order}
In this appendix, we shall discuss the subleading corrections to the
vacuum alignment induced by the higher dimensional operators. At the
next level of approximation, the driving superpotential $w_v$ in
Eq.(\ref{24}) is modified into $w_v+\delta w_v$. Due to the
constraint imposed by $Z_4\times Z_2$ symmetry, the correction terms
are suppressed by $1/\Lambda^2$. Concretely $\delta w_v$ is given by
\begin{equation}
\label{b1}\delta w_v=\frac{1}{\Lambda^2}\sum^{34}_{i=1}c_i{\cal
O}^{\chi^0}_i+\frac{1}{\Lambda^2}\sum^{11}_{i=1}r_i{\cal
O}^{\rho^{0}}_i+\frac{1}{\Lambda^2}\sum^{11}_{i=1}t_i{\cal
O}^{\theta^{0}}_i+\frac{1}{\Lambda^2}\sum^{28}_{i=1}e_i{\cal
O}^{\eta^{0}}_i+\frac{1}{\Lambda^2}\sum^{19}_{i=1}k_i{\cal
O}^{\xi^0}_i
\end{equation}
where $c_i$, $r_i$, $t_i$, $e_i$ and $k_i$ are order one
coefficients, $\{{\cal O}^{\chi^{0}}_i, {\cal O}^{\rho^0}_i, {\cal
O}^{\theta^0}_i, {\cal O}^{\eta^{0}}_i, {\cal O}^{\xi^0}_i\}$ denote
the complete set of subleading contractions invariant under
$T_{13}\times Z_4\times Z_2$.
\begin{eqnarray}
\nonumber&&{\cal
O}^{\chi^0}_1=\chi^0((\chi\chi)_{\mathbf{3}_{1S}}(\phi\phi)_{\mathbf{\bar{3}_2}})_{\mathbf{\bar{3}_2}},~~{\cal
O}^{\chi^0}_2=\chi^0((\chi\chi)_{\mathbf{3}_{1S}}(\eta\eta)_{\mathbf{3}_{2S}})_{\mathbf{\bar{3}_2}},~~{\cal
O}^{\chi^0}_3=\chi^0((\chi\chi)_{\mathbf{3}_{1S}}(\phi\eta)_{\mathbf{\bar{3}_1}})_{\mathbf{\bar{3}_2}}\\
\nonumber&&{\cal
O}^{\chi^0}_4=\chi^0((\chi\chi)_{\mathbf{3}_{1S}}(\phi\eta)_{\mathbf{3_2}})_{\mathbf{\bar{3}_2}},~~{\cal
O}^{\chi^0}_5=\chi^0((\chi\chi)_{\mathbf{3}_{1S}}(\phi\eta)_{\mathbf{\bar{3}_2}})_{\mathbf{\bar{3}_2}},~~{\cal
O}^{\chi^0}_6=\chi^0((\chi\chi)_{\mathbf{\bar{3}_2}}(\phi\phi)_{\mathbf{3}_{1S}})_{\mathbf{\bar{3}_2}},\\
\nonumber&&{\cal
O}^{\chi^0}_7=\chi^0((\chi\chi)_{\mathbf{\bar{3}_2}}(\eta\eta)_{\mathbf{3_1}})_{\mathbf{\bar{3}_2}},~~{\cal
O}^{\chi^0}_8=\chi^0((\chi\chi)_{\mathbf{\bar{3}_2}}(\phi\eta)_{\mathbf{\bar{3}_1}})_{\mathbf{\bar{3}_2}},~~{\cal
O}^{\chi^0}_9=\chi^0((\xi\xi)_{\mathbf{\bar{3}}_{1S}}(\phi\phi)_{\mathbf{3}_{1S}})_{\mathbf{\bar{3}_2}},\\
\nonumber&&{\cal
O}^{\chi^0}_{10}=\chi^0((\xi\xi)_{\mathbf{\bar{3}}_{1S}}(\phi\phi)_{\mathbf{\bar{3}_2}})_{\mathbf{\bar{3}_2}},~~{\cal
O}^{\chi^0}_{11}=\chi^0((\xi\xi)_{\mathbf{\bar{3}}_{1S}}(\eta\eta)_{\mathbf{3_1}})_{\mathbf{\bar{3}_2}},~~{\cal
O}^{\chi^0}_{12}=\chi^0((\xi\xi)_{\mathbf{\bar{3}}_{1S}}(\phi\eta)_{\mathbf{\bar{3}_1}})_{\mathbf{\bar{3}_2}},\\
\nonumber&&{\cal
O}^{\chi^0}_{13}=\chi^0((\xi\xi)_{\mathbf{\bar{3}}_{1S}}(\phi\eta)_{\mathbf{\bar{3}_2}})_{\mathbf{\bar{3}_2}},~~{\cal
O}^{\chi^0}_{14}=\chi^0((\xi\xi)_{\mathbf{3_2}}(\phi\phi)_{\mathbf{3}_{1S}})_{\mathbf{\bar{3}}_2},~~{\cal
O}^{\chi^0}_{15}=\chi^0((\xi\xi)_{\mathbf{3_2}}(\eta\eta)_{\mathbf{3_1}})_{\mathbf{\bar{3}_2}},\\
\nonumber&&{\cal
O}^{\chi^0}_{16}=\chi^0((\xi\xi)_{\mathbf{3_2}}(\eta\eta)_{\mathbf{3}_{2S}})_{\mathbf{\bar{3}}_{2S}},~~{\cal
O}^{\chi^0}_{17}=\chi^0((\xi\xi)_{\mathbf{3_2}}(\eta\eta)_{\mathbf{3}_{2S}})_{\mathbf{\bar{3}}_{2A}},~~{\cal
O}^{\chi^0}_{18}=\chi^0((\xi\xi)_{\mathbf{3_2}}(\phi\eta)_{\mathbf{3_2}})_{\mathbf{\bar{3}}_{2S}},\\
\nonumber&&{\cal
O}^{\chi^0}_{19}=\chi^0((\xi\xi)_{\mathbf{3_2}}(\phi\eta)_{\mathbf{3_2}})_{\mathbf{\bar{3}}_{2A}},~~{\cal
O}^{\chi^0}_{20}=\chi^0(\chi\xi)_{\mathbf{1_1}}(\phi\phi)_{\mathbf{\bar{3}_2}},~~{\cal
O}^{\chi^0}_{21}=\chi^0(\chi\xi)_{\mathbf{1_1}}(\phi\eta)_{\mathbf{\bar{3}_2}},\\
\nonumber&&{\cal
O}^{\chi^0}_{22}=\chi^0(\chi\xi)_{\mathbf{1_2}}(\phi\phi)_{\mathbf{\bar{3}_2}},~~{\cal
O}^{\chi^0}_{23}=\chi^0(\chi\xi)_{\mathbf{1_2}}(\phi\eta)_{\mathbf{\bar{3}_2}},~~{\cal
O}^{\chi^0}_{24}=\chi^0(\chi\xi)_{\mathbf{1_3}}(\phi\phi)_{\mathbf{\bar{3}_2}},\\
\nonumber&&{\cal
O}^{\chi^0}_{25}=\chi^0(\chi\xi)_{\mathbf{1_3}}(\phi\eta)_{\mathbf{\bar{3}_2}},~~{\cal
O}^{\chi^0}_{26}=\chi^0((\chi\xi)_{\mathbf{3_2}}(\phi\phi)_{\mathbf{3}_{1S}})_{\mathbf{\bar{3}_2}},~~{\cal
O}^{\chi^0}_{27}=\chi^0((\chi\xi)_{\mathbf{3_2}}(\eta\eta)_{\mathbf{3_1}})_{\mathbf{\bar{3}_2}},\\
\nonumber&&{\cal
O}^{\chi^0}_{28}=\chi^0((\chi\xi)_{\mathbf{3_2}}(\eta\eta)_{\mathbf{3}_{2S}})_{\mathbf{\bar{3}}_{2S}},~~{\cal
O}^{\chi^0}_{29}=\chi^0((\chi\xi)_{\mathbf{3_2}}(\eta\eta)_{\mathbf{3}_{2S}})_{\mathbf{\bar{3}}_{2A}},~~{\cal
O}^{\chi^0}_{30}=\chi^0((\chi\xi)_{\mathbf{3_2}}(\phi\eta)_{\mathbf{3_2}})_{\mathbf{\bar{3}}_{2S}},\\
\nonumber&&{\cal
O}^{\chi^0}_{31}=\chi^0((\chi\xi)_{\mathbf{3_2}}(\phi\eta)_{\mathbf{3_2}})_{\mathbf{\bar{3}}_{2A}},~~{\cal
O}^{\chi^0}_{32}=\chi^0((\chi\xi)_{\mathbf{\bar{3}_2}}(\phi\phi)_{\mathbf{3}_{1S}})_{\mathbf{\bar{3}_2}},~~{\cal
O}^{\chi^0}_{33}=\chi^0((\chi\xi)_{\mathbf{\bar{3}_2}}(\eta\eta)_{\mathbf{3_1}})_{\mathbf{\bar{3}_2}}\\
\label{b2}&&{\cal
O}^{\chi^0}_{34}=\chi^0((\chi\xi)_{\mathbf{\bar{3}_2}}(\phi\eta)_{\mathbf{\bar{3}_1}})_{\mathbf{\bar{3}_2}}
\end{eqnarray}
\begin{eqnarray}
\nonumber&&{\cal
O}^{\rho^{0}}_1=\rho^{0}((\chi\chi)_{\mathbf{3}_{1S}}(\phi\eta)_{\mathbf{\bar{3}_1}})_{\mathbf{1_3}},~~{\cal
O}^{\rho^{0}}_2=\rho^{0}((\chi\chi)_{\mathbf{\bar{3}_2}}(\eta\eta)_{\mathbf{3}_{2S}})_{\mathbf{1_3}},~~{\cal
O}^{\rho^{0}}_3=\rho^{0}((\chi\chi)_{\mathbf{\bar{3}_2}}(\phi\eta)_{\mathbf{3_2}})_{\mathbf{1_3}},\\
\nonumber&&{\cal
O}^{\rho^{0}}_4=\rho^{0}((\xi\xi)_{\mathbf{\bar{3}}_{1S}}(\phi\phi)_{\mathbf{3}_{1S}})_{\mathbf{1_3}},~~{\cal
O}^{\rho^{0}}_5=\rho^{0}((\xi\xi)_{\mathbf{\bar{3}}_{1S}}(\eta\eta)_{\mathbf{3_1}})_{\mathbf{1_3}},~~{\cal
O}^{\rho^{0}}_6=\rho^{0}((\xi\xi)_{\mathbf{3_2}}(\phi\phi)_{\mathbf{\bar{3}_2}})_{\mathbf{1_3}},\\
\nonumber&&{\cal
O}^{\rho^{0}}_7=\rho^{0}((\xi\xi)_{\mathbf{3_2}}(\phi\eta)_{\mathbf{\bar{3}_2}})_{\mathbf{1_3}},~~{\cal
O}^{\rho^{0}}_8=\rho^{0}((\chi\xi)_{\mathbf{3_2}}(\phi\phi)_{\mathbf{\bar{3}_2}})_{\mathbf{1_3}},~~{\cal
O}^{\rho^{0}}_9=\rho^{0}((\chi\xi)_{\mathbf{3_2}}(\phi\eta)_{\mathbf{\bar{3}_2}})_{\mathbf{1_3}},\\
\label{b5}&&{\cal
O}^{\rho^{0}}_{10}=\rho^{0}((\chi\xi)_{\mathbf{\bar{3}_2}}(\eta\eta)_{\mathbf{3}_{2S}})_{\mathbf{1_3}},~~{\cal
O}^{\rho^{0}}_{11}=\rho^{0}((\chi\xi)_{\mathbf{\bar{3}_2}}(\phi\eta)_{\mathbf{3_2}})_{\mathbf{1_3}}
\end{eqnarray}
\begin{eqnarray}
\nonumber&&{\cal
O}^{\theta^{0}}_1=\theta^{0}((\chi\chi)_{\mathbf{3}_{1S}}(\phi\eta)_{\mathbf{\bar{3}_1}})_{\mathbf{1_2}},~~{\cal
O}^{\theta^{0}}_2=\theta^{0}((\chi\chi)_{\mathbf{\bar{3}_2}}(\eta\eta)_{\mathbf{3}_{2S}})_{\mathbf{1_2}},~~{\cal
O}^{\theta^{0}}_3=\theta^{0}((\chi\chi)_{\mathbf{\bar{3}_2}}(\phi\eta)_{\mathbf{3_2}})_{\mathbf{1_2}},\\
\nonumber&&{\cal
O}^{\theta^{0}}_4=\theta^{0}((\xi\xi)_{\mathbf{\bar{3}}_{1S}}(\phi\phi)_{\mathbf{3}_{1S}})_{\mathbf{1_2}},~~{\cal
O}^{\theta^{0}}_5=\theta^{0}((\xi\xi)_{\mathbf{\bar{3}}_{1S}}(\eta\eta)_{\mathbf{3_1}})_{\mathbf{1_2}},~~{\cal
O}^{\theta^{0}}_6=\theta^{0}((\xi\xi)_{\mathbf{3_2}}(\phi\phi)_{\mathbf{\bar{3}_2}})_{\mathbf{1_2}},\\
\nonumber&&{\cal
O}^{\theta^{0}}_7=\theta^{0}((\xi\xi)_{\mathbf{3_2}}(\phi\eta)_{\mathbf{\bar{3}_2}})_{\mathbf{1_2}},~~{\cal
O}^{\theta^{0}}_8=\theta^{0}((\chi\xi)_{\mathbf{3_2}}(\phi\phi)_{\mathbf{\bar{3}_2}})_{\mathbf{1_2}},~~{\cal
O}^{\theta^{0}}_9=\theta^{0}((\chi\xi)_{\mathbf{3_2}}(\phi\eta)_{\mathbf{\bar{3}_2}})_{\mathbf{1_2}},\\
\label{b6}&&{\cal
O}^{\theta^{0}}_{10}=\theta^{0}((\chi\xi)_{\mathbf{\bar{3}_2}}(\eta\eta)_{\mathbf{3}_{2S}})_{\mathbf{1_2}},~~{\cal
O}^{\theta^{0}}_{11}=\theta^{0}((\chi\xi)_{\mathbf{\bar{3}_2}}(\phi\eta)_{\mathbf{3_2}})_{\mathbf{1_2}}
\end{eqnarray}
\begin{eqnarray}
\nonumber&&{\cal
O}^{\eta^0}_{1}=\eta^0((\phi\phi)_{\mathbf{3}_{1S}}(\phi\phi)_{\mathbf{3}_{1S}})_{\mathbf{3_2}},~~{\cal
O}^{\eta^0}_{2}=\eta^0((\phi\phi)_{\mathbf{\bar{3}_2}}(\phi\phi)_{\mathbf{\bar{3}_2}})_{\mathbf{3}_{2S}},~~{\cal
O}^{\eta^0}_{3}=\eta^0((\phi\phi)_{\mathbf{3}_{1S}}(\phi\eta)_{\mathbf{\bar{3}_1}})_{\mathbf{3_2}},\\
\nonumber&&{\cal
O}^{\eta^0}_{4}=\eta^0((\phi\phi)_{\mathbf{3}_{1S}}(\phi\eta)_{\mathbf{3_2}})_{\mathbf{3_2}},~~{\cal
O}^{\eta^0}_{5}=\eta^0((\phi\phi)_{\mathbf{\bar{3}_2}}(\phi\eta)_{\mathbf{\bar{3}_1}})_{\mathbf{3_2}},~~{\cal
O}^{\eta^0}_{6}=\eta^0((\phi\phi)_{\mathbf{\bar{3}_2}}(\phi\eta)_{\mathbf{\bar{3}_2}})_{\mathbf{3}_{2S}},\\
\nonumber&&{\cal
O}^{\eta^0}_{7}=\eta^0((\phi\phi)_{\mathbf{\bar{3}_2}}(\phi\eta)_{\mathbf{\bar{3}_2}})_{\mathbf{3}_{2A}},~~{\cal
O}^{\eta^0}_{8}=\eta^0((\phi\phi)_{\mathbf{3}_{1S}}(\eta\eta)_{\mathbf{3_1}})_{\mathbf{3_2}},~~{\cal
O}^{\eta^0}_{9}=\eta^0((\phi\phi)_{\mathbf{3}_{1S}}(\eta\eta)_{\mathbf{3}_{2S}})_{\mathbf{3_2}},\\
\nonumber&&{\cal
O}^{\eta^0}_{10}=\eta^0((\eta\eta)_{\mathbf{3_1}}(\phi\eta)_{\mathbf{\bar{3}_1}})_{\mathbf{3_2}},~~{\cal
O}^{\eta^0}_{11}=\eta^0((\eta\eta)_{\mathbf{3_1}}(\phi\eta)_{\mathbf{3_2}})_{\mathbf{3_2}},~~{\cal
O}^{\eta^0}_{12}=\eta^0((\eta\eta)_{\mathbf{3}_{2S}}(\phi\eta)_{\mathbf{\bar{3}_1}})_{\mathbf{3_2}},\\
\nonumber&&{\cal
O}^{\eta^0}_{13}=\eta^0((\eta\eta)_{\mathbf{3_1}}(\eta\eta)_{\mathbf{3_1}})_{\mathbf{3_2}},~~{\cal
O}^{\eta^0}_{14}=\eta^0((\eta\eta)_{\mathbf{3_1}}(\eta\eta)_{\mathbf{3}_{2S}})_{\mathbf{3_2}},~~{\cal
O}^{\eta^0}_{15}=\eta^0((\chi\chi)_{\mathbf{3}_{1S}}(\chi\chi)_{\mathbf{3}_{1S}})_{\mathbf{3_2}},\\
\nonumber&&{\cal
O}^{\eta^0}_{16}=\eta^0((\chi\chi)_{\mathbf{\bar{3}_2}}(\chi\chi)_{\mathbf{\bar{3}_2}})_{\mathbf{3}_{2S}},~~{\cal
O}^{\eta^0}_{17}=\eta^0((\chi\chi)_{\mathbf{3}_{1S}}(\chi\xi)_{\mathbf{3_2}})_{\mathbf{3_2}},~~{\cal
O}^{\eta^0}_{18}=\eta^0((\chi\chi)_{\mathbf{\bar{3}_2}}(\chi\xi)_{\mathbf{\bar{3}_2}})_{\mathbf{3}_{2S}},\\
\nonumber&&{\cal
O}^{\eta^0}_{19}=\eta^0((\chi\chi)_{\mathbf{\bar{3}_2}}(\chi\xi)_{\mathbf{\bar{3}_2}})_{\mathbf{3}_{2A}},~~{\cal
O}^{\eta^0}_{20}=\eta^0((\chi\chi)_{\mathbf{3}_{1S}}(\xi\xi)_{\mathbf{\bar{3}}_{1S}})_{\mathbf{3_2}},~~{\cal
O}^{\eta^0}_{21}=\eta^0((\chi\chi)_{\mathbf{3}_{1S}}(\xi\xi)_{\mathbf{3_2}})_{\mathbf{3_2}},\\
\nonumber&&{\cal
O}^{\eta^0}_{22}=\eta^0((\chi\chi)_{\mathbf{\bar{3}_2}}(\xi\xi)_{\mathbf{\bar{3}}_{1S}})_{\mathbf{3_2}},~~{\cal
O}^{\eta^0}_{23}=\eta^0((\xi\xi)_{\mathbf{\bar{3}}_{1S}}(\chi\xi)_{\mathbf{3_2}})_{\mathbf{3_2}},~~{\cal
O}^{\eta^0}_{24}=\eta^0((\xi\xi)_{\mathbf{\bar{3}}_{1S}}(\chi\xi)_{\mathbf{\bar{3}_2}})_{\mathbf{3_2}},\\
\nonumber&&{\cal
O}^{\eta^0}_{25}=\eta^0(\xi\xi)_{\mathbf{3_2}}(\chi\xi)_{1_1},~~{\cal
O}^{\eta^0}_{26}=\eta^0(\xi\xi)_{\mathbf{3_2}}(\chi\xi)_{\mathbf{1_2}},~~{\cal
O}^{\eta^0}_{27}=\eta^0(\xi\xi)_{\mathbf{3_2}}(\chi\xi)_{\mathbf{1_3}},\\
\label{b3}&&{\cal
O}^{\eta^0}_{28}=\eta^0((\xi\xi)_{\mathbf{\bar{3}}_{1S}}(\xi\xi)_{\mathbf{3_2}})_{\mathbf{3_2}}
\end{eqnarray}
\begin{eqnarray}
\nonumber&&{\cal
O}^{\xi^0}_1=\xi^{0}((\chi\phi)_{\mathbf{3}_{1S}}(\phi\eta)_{\mathbf{\bar{3}_1}})_{\mathbf{1_1}},~~{\cal
O}^{\xi^0}_2=\xi^{0}((\chi\phi)_{\mathbf{3}_{1A}}(\phi\eta)_{\mathbf{\bar{3}_1}})_{\mathbf{1_1}},~~{\cal
O}^{\xi^0}_3=\xi^{0}((\chi\phi)_{\mathbf{\bar{3}_2}}(\eta\eta)_{\mathbf{3}_{2S}})_{\mathbf{1_1}},\\
\nonumber&&{\cal
O}^{\xi^0}_4=\xi^{0}((\chi\phi)_{\mathbf{\bar{3}_2}}(\phi\eta)_{\mathbf{3_2}})_{\mathbf{1_1}},~~{\cal
O}^{\xi^0}_5=\xi^{0}((\chi\eta)_{\mathbf{\bar{3}_1}}(\eta\eta)_{\mathbf{3_1}})_{\mathbf{1_1}},~~{\cal
O}^{\xi^0}_6=\xi^{0}((\chi\eta)_{\mathbf{\bar{3}_2}}(\eta\eta)_{\mathbf{3}_{2S}})_{\mathbf{1_1}},\\
\nonumber&&{\cal
O}^{\xi^0}_7=\xi^{0}((\xi\phi)_{\mathbf{3_2}}(\phi\phi)_{\mathbf{\bar{3}_2}})_{\mathbf{1_1}},~~{\cal
O}^{\xi^0}_8=\xi^{0}((\xi\phi)_{\mathbf{3_2}}(\phi\eta)_{\mathbf{\bar{3}_2}})_{\mathbf{1_1}},~~{\cal
O}^{\xi^0}_9=\xi^{0}((\xi\phi)_{\mathbf{\bar{3}_2}}(\eta\eta)_{\mathbf{3}_{2S}})_{\mathbf{1_1}},\\
\label{b4}&&{\cal
O}^{\xi^0}_{10}=\xi^{0}((\xi\phi)_{\mathbf{\bar{3}_2}}(\phi\eta)_{\mathbf{3_2}})_{\mathbf{1_1}},~~{\cal
O}^{\xi^0}_{11}=\xi^{0}((\xi\eta)_{\mathbf{\bar{3}_1}}(\eta\eta)_{\mathbf{3_1}})_{\mathbf{1_1}},~~{\cal
O}^{\xi^0}_{12}=\xi^{0}((\xi\eta)_{\mathbf{\bar{3}_2}}(\eta\eta)_{\mathbf{3}_{2S}})_{\mathbf{1_1}}
\end{eqnarray}
The subleading contribution $\delta w_v$ modifies the LO VEVs, then
the new vacuum configuration can be parameterized as
\begin{eqnarray}
\nonumber&&\langle\chi\rangle=\left(\begin{array}{c} v_{\chi}+\delta
v_{\chi_1}\\
v_{\chi}+\delta
v_{\chi_2}\\
v_{\chi}+\delta v_{\chi_3}
\end{array}\right),~~~~~~~\langle\xi\rangle=\left(\begin{array}{c}v_{\xi}+\delta v_{\xi_1}\\
v_{\xi}+\delta v_{\xi_2}\\
v_{\xi}
\end{array}\right),\\
\label{b7}&&\langle\phi\rangle=\left(\begin{array}{c} \delta
v_{\phi_1}\\
v_{\phi}+\delta v_{\phi_2}\\
-v_{\phi}
\end{array}\right),~~~~~~~\langle\eta\rangle=\left(\begin{array}{c}\delta v_{\eta_1}\\
v_{\eta}\\
\delta v_{\eta_3}
\end{array}\right)
\end{eqnarray}
where the shifts $\delta v_{\xi_3}$, $\delta v_{\phi_3}$ and $\delta
v_{\eta_2}$ have been absorbed into the undetermined parameters
$v_{\xi}$, $v_{\phi}$ and $v_{\eta}$. Similar to section 4, the new
vacua is obtained by searching for the zeros of the F-terms, i.e.
the first derivative of $w_v+\delta w_v$ with respect to the driving
fields $\chi^0$, $\rho^0$, $\theta^{0}$, $\eta^{0}$ and $\xi^{0}$.
By keeping only the terms linear in the shift $\delta v$ and
neglecting the terms proportional to $\delta v/\Lambda$, the
minimization equations become
\begin{eqnarray}
\nonumber&&2f_1v_{\chi}\delta v_{\chi_1}+f_2v_{\xi}\delta
v_{\chi_2}+f_2v_{\chi}\delta
v_{\xi_1}+a_1v_{\chi}v_{\xi}v^2_{\phi}/\Lambda^2=0\\
\nonumber&&2f_1v_{\chi}\delta v_{\chi_2}+f_2v_{\xi}\delta
v_{\chi_3}+f_2v_{\chi}\delta
v_{\xi_2}+a_2v_{\chi}v_{\xi}v^2_{\phi}/\Lambda^2=0\\
\nonumber&&2f_1v_{\chi}\delta v_{\chi_3}+f_2v_{\xi}\delta
v_{\chi_1}+a_3v_{\chi}v_{\xi}v^2_{\phi}/\Lambda^2=0\\
\nonumber&&f_3[v_{\xi}(\delta v_{\chi_1}+\omega^2\delta
v_{\chi_2}+\omega\delta v_{\chi_3})+v_{\chi}(\delta
v_{\xi_1}+\omega^2\delta
v_{\xi_2})]+a_4v_{\chi}v_{\xi}v^2_{\phi}/\Lambda^2=0\\
\label{b8}&&f_4[v_{\xi}(\delta v_{\chi_1}+\omega\delta
v_{\chi_2}+\omega^2\delta v_{\chi_3})+v_{\chi}(\delta
v_{\xi_1}+\omega\delta
v_{\xi_2})]+a_5v_{\chi}v_{\xi}v^2_{\phi}/\Lambda^2=0
\end{eqnarray}
where the coefficients $a_i(i=1-5)$ are linear combinations of the
subleading coefficients
\begin{eqnarray}
\nonumber&&a_1=2c_1v_{\chi}/v_{\xi}+c_8v_{\chi}v_{\eta}/(v_{\xi}v_{\phi})+(-4c_9+2c_{10})v_{\xi}/v_{\chi}+2c_{11}v_{\xi}v^2_{\eta}/(v_{\chi}v^2_{\phi})
+(-3c_{21}+c_{34})v_{\eta}/v_{\phi}\\
\nonumber&&a_2=2(c_3-c_5)v_{\chi}v_{\eta}/(v_{\xi}v_{\phi})+2(c_{10}-c_{14})v_{\xi}/v_{\phi}+c_{15}v_{\xi}v^2_{\eta}/(v_{\chi}v^2_{\phi})
+3(c_{20}-c_{26})+c_{27}v^2_{\eta}/v^2_{\phi}\\
\nonumber&&a_3=2(c_1-c_6)v_{\chi}/v_{\xi}+c_7v_{\chi}v^2_{\eta}/(v_{\xi}v^2_{\phi})+2(c_{12}-c_{13})v_{\xi}v_{\eta}/(v_{\chi}v_{\phi})
+3c_{20}-2c_{32}+c_{33}v^2_{\eta}/v^2_{\phi}\\
\nonumber&&a_4=2\omega
r_1v_{\chi}v_{\eta}/(v_{\xi}v_{\phi})-(4r_4+r_6)v_{\xi}/v_{\chi}+2r_5v_{\xi}v^2_{\eta}/(v_{\chi}v^2_{\phi})-r_7v_{\xi}v_{\eta}/(v_{\chi}v_{\phi})
-r_8-r_9v_{\eta}/v_{\phi}\\
\label{b9}&&a_5=2\omega^2
t_1v_{\chi}v_{\eta}/(v_{\xi}v_{\phi})-(4t_4+t_6)v_{\xi}/v_{\chi}+2t_5v_{\xi}v^2_{\eta}/(v_{\chi}v^2_{\phi})-t_7v_{\xi}v_{\eta}/(v_{\chi}v_{\phi})
-t_8-t_9v_{\eta}/v_{\phi}
\end{eqnarray}
The equations Eq.(\ref{b8}) are linear in $\delta v_{\chi_i}(i=1, 2,
3)$ and $\delta v_{\xi_i}(i=1,2)$, and can be solved
straightforwardly by
\begin{eqnarray}
\nonumber&&\frac{\delta
v_{\chi_1}}{v_{\chi}}=(6a_1+2a_2+5a_3)\frac{v^2_{\phi}}{13f_2\Lambda^2}-2(7+5\omega)a_4\frac{v^2_{\phi}}{39f_3\Lambda^2}-2(2-5\omega)a_5\frac{v^2_{\phi}}{39f_4\Lambda^2}\\
\nonumber&&\frac{\delta
v_{\chi_2}}{v_{\chi}}=(2a_1+5a_2+6a_3)\frac{v^2_{\phi}}{13f_2\Lambda^2}-(3+4\omega)a_4\frac{v^2_{\phi}}{13f_3\Lambda^2}
+(1+4\omega)a_5\frac{v^2_{\phi}}{13f_4\Lambda^2}\\
\nonumber&&\frac{\delta
v_{\chi_3}}{v_{\chi}}=(3a_1+a_2+9a_3)\frac{v^2_{\phi}}{13f_2\Lambda^2}-(7+5\omega)a_4\frac{v^2_{\phi}}{39f_3\Lambda^2}
-(2-5\omega)a_5\frac{v^2_{\phi}}{39f_4\Lambda^2}\\
\nonumber&&\frac{\delta
v_{\xi_1}}{v_{\xi}}=-(3a_1+a_2-4a_3)\frac{v^2_{\phi}}{13f_2\Lambda^2}-(19+8\omega)a_4\frac{v^2_{\phi}}{39f_3\Lambda^2}
-(11-8\omega)a_5\frac{v^2_{\phi}}{39f_4\Lambda^2}\\
\label{b10}&&\frac{\delta
v_{\xi_2}}{v_{\xi}}=(a_1-4a_2+3a_3)\frac{v^2_{\phi}}{13f_2\Lambda^2}-(11+19\omega)a_4\frac{v^2_{\phi}}{39f_3\Lambda^2}
+(8+19\omega)a_5\frac{v^2_{\phi}}{39f_4\Lambda^2}
\end{eqnarray}
From the above equations, we clearly see that all the shifts $\delta
v_{\chi_1}/v_{\chi}$, $\delta v_{\chi_2}/v_{\chi}$, $\delta
v_{\chi_3}/v_{\chi}$, $\delta v_{\xi_1}/v_{\xi}$ and $\delta
v_{\xi_2}/v_{\xi}$ are of order $\varepsilon'^{2}$. The minimization
equations for $\delta v_{\phi_1}$, $\delta v_{\phi_2}$, $\delta
v_{\eta_1}$ and $\delta v_{\eta_3}$ are
\begin{eqnarray}
\nonumber&&2g_1v_{\eta}\delta v_{\eta_3}-g_2v_{\phi}\delta
v_{\eta_1}+b_1v_{\eta}v^3_{\phi}/\Lambda^2=0\\
\nonumber&&g_2v_{\eta}\delta
v_{\phi_1}+b_2v_{\eta}v^3_{\phi}/\Lambda^2=0\\
\nonumber&&2g_1v_{\eta}\delta v_{\eta_1}+g_2v_{\phi}\delta
v_{\eta_3}+b_3v_{\eta}v^3_{\phi}/\Lambda^2=0\\
\label{b11}&&h(v_{\xi}\delta v_{\phi_1}+v_{\xi}\delta
v_{\phi_2}+v_{\phi}\delta
v_{\xi_2})+b_4v_{\xi}v^3_{\phi}/\Lambda^2=0
\end{eqnarray}
where the coefficients $b_i(i=1-4)$ are given by
\begin{eqnarray}
\nonumber&&d=[2(2e_{15}+e_{16})v^4_{\chi}+2(e_{17}+e_{18})v^3_{\chi}v_{\xi}+2(2e_{20}+e_{21}+e_{22})v^2_{\chi}v^2_{\xi}+(2e_{23}+2e_{24}
+3e_{25})v_{\chi}v^3_{\xi}\\
\nonumber&&~~+2e_{28}v^4_{\xi}]/(v_{\eta}v^3_{\phi})\\  
\nonumber&&b_1=(4e_1+2e_2)v_{\phi}/v_{\eta}-2e_8v_{\eta}/v_{\phi}+e_{13}v^3_{\eta}/v^3_{\phi}+d\\
\nonumber&&b_2=-e_6-e_7+d\\
\nonumber&&b_3=-2e_3-e_6+e_7+e_{10}v^2_{\eta}/v^2_{\phi}+d\\
\label{b12}&&b_4=(k_1+k_2)v_{\chi}v_{\eta}/(v_{\xi}v_{\phi})
\end{eqnarray}
The solutions to Eq.(\ref{b11}) are given by
\begin{eqnarray}
\nonumber&&\frac{\delta
v_{\phi_1}}{v_{\phi}}=-\frac{b_2}{g_2}\frac{v^2_{\phi}}{\Lambda^2}\\
\nonumber&&\frac{\delta
v_{\phi_2}}{v_{\phi}}=(\frac{b_2}{g_2}-\frac{b_4}{h})\frac{v^2_{\phi}}{\Lambda^2}-\frac{\delta
v_{\xi_2}}{v_{\xi}}\\
\nonumber&&\frac{\delta
v_{\eta_1}}{v_{\eta}}=\frac{(b_1g_2v_{\phi}-2b_3g_1v_{\eta})v_{\phi}}{4g^2_1v^2_{\eta}+g^2_2v^2_{\phi}}\frac{v^2_{\phi}}{\Lambda^2}\\
\label{b13}&&\frac{\delta
v_{\eta_3}}{v_{\eta}}=-\frac{(2b_1g_1v_{\eta}+b_3g_2v_{\phi})v_{\phi}}{4g^2_1v^2_{\eta}+g^2_2v^2_{\phi}}\frac{v^2_{\phi}}{\Lambda^2}
\end{eqnarray}
Obviously $\delta v_{\phi_1}/v_{\phi}$, $\delta
v_{\phi_2}/v_{\phi}$, $\delta v_{\eta_1}/v_{\eta}$ and $\delta
v_{\eta_3}/v_{\eta}$ are of order $\varepsilon'^2$ as well. As is
shown in Eq.(\ref{b3}), the subleading terms proportional to
$\eta^{0}$ are of the structures $\eta^0\Phi^4_{\nu}$ or
$\eta^{0}\Phi^4_{\ell}$, the contributions of the latter operator to
the vacuum alignment are parameterized in terms of the parameter $d$
in Eq.(\ref{b12}). If we have a large VEV of $\Phi_{\nu}$ with
$\langle\Phi_{\nu}\rangle/\Lambda\sim\lambda_c$, then the structure
$\eta^0\Phi^4_{\nu}$ is dominant. On the other hand, if the VEVs of
$\Phi_{\nu}$ and $\Phi_{\ell}$ are of the same order of magnitude,
the contributions of the two type of operators are comparable.

\end{document}